\documentclass[english]{emulateapj}

\def\gs{\mathrel{\raise1.16pt\hbox{$>$}\kern-7.0pt %
\lower3.06pt\hbox{{$\scriptstyle \sim$}}}}         %
\def\ls{\mathrel{\raise1.16pt\hbox{$<$}\kern-7.0pt %
\lower3.06pt\hbox{{$\scriptstyle \sim$}}}}         %
 \newcommand{\ltsimeq}{\raisebox{-0.6ex}{$\,\stackrel 
        {\raisebox{-.2ex}{$\textstyle <$}}{\sim}\,$}} 
\newcommand{\gtsimeq}{\raisebox{-0.6ex}{$\,\stackrel 
        {\raisebox{-.2ex}{$\textstyle >$}}{\sim}\,$}}

\slugcomment{}
\shorttitle{Spitzer IRS Spectra of X-ray Sources}
\shortauthors{Brand et al.}

\begin{document}

\title{$Spitzer$ Mid-Infrared Spectroscopy of Distant X-ray Luminous AGN}

\author{Kate Brand\altaffilmark{1}, Dan ~W. Weedman\altaffilmark{2}, Vandana Desai\altaffilmark{3}, Emeric Le Floc'h\altaffilmark{4}, Lee Armus\altaffilmark{5}, Arjun Dey\altaffilmark{6}, Jim ~R. Houck\altaffilmark{2}, Buell ~T. Jannuzi\altaffilmark{6}, Howard ~A. Smith\altaffilmark{7}, B.~T. Soifer\altaffilmark{3,5}} 

\altaffiltext{1}{Space Telescope Science Institute, 3700 San Martin Drive, Baltimore, MD 21218; brand@stsci.edu}
\altaffiltext{2}{Astronomy Department, Cornell University, Ithica, NY 14853}
\altaffiltext{3}{Division of Physics, Mathematics and Astronomy, California Institute of Technology, 320-47, Pasadena, CA 91125}
\altaffiltext{4}{Institute for Astronomy, University of Hawaii, 2680 Woodlawn Drive, Honolulu, HI 96822, USA}
\altaffiltext{5}{Spitzer Science Center, California Institute of Technology, 220-6, Pasadena, CA 91125}
\altaffiltext{6}{National Optical Astronomy Observatory, 950 North Cherry Avenue, Tucson, AZ 85726} 
\altaffiltext{7}{Harvard-Smithsonian Center for Astrophysics, 60 Garden Street, Cambridge, MA 02138}

\begin{abstract}

We present mid-infrared spectroscopy of a sample of 16 optically faint infrared luminous galaxies obtained with the Infrared Spectrograph (IRS) on the $Spitzer~Space~Telescope$. These sources were jointly selected from $Spitzer$ and $Chandra$ imaging surveys in the NDWFS Bo\"otes field and were selected from their bright X-ray fluxes to host luminous AGN. None of the spectra show significant emission from polycyclic aromatic hydrocarbons (PAHs; 6.2$\rm \mu m$ equivalent widths $<$0.2$\rm \mu m$), consistent with their infrared emission being dominated by AGN. Nine of the X-ray sources show 9.7$\rm \mu m$ silicate absorption features. Their redshifts are in the range $0.9<z<2.6$, implying infrared luminosities of log(L$_{IR}$)=12.5$-$13.6 $L_\odot$. The average silicate absorption strength is not as strong as that of previously targeted optically faint infrared luminous galaxies with similar mid-infrared luminosities implying that the X-ray selection favors sources behind a smaller column of Si-rich dust than non-X-ray selection. Seven of the X-ray sources have featureless power-law mid-IR spectra. We argue that the featureless spectra likely result from the sources having weak or absent silicate and PAH features rather than the sources lying at higher redshifts where these features are shifted out of the IRS spectral window. We investigate whether there are any correlations between X-ray and infrared properties and find that sources with silicate absorption features tend to have fainter X-ray fluxes and harder X-ray spectra, indicating a weak relation between the amount of silicate absorption and column density of X-ray-absorbing gas. 
\end{abstract}


\section{Introduction}

Mid-infrared spectroscopy is a useful tool in investigating the extent to which starbursts and AGN contribute to the luminosity of luminous infrared galaxies (LIRGs; \citealt{san88}). The infrared spectra of local LIRGs have been studied extensively with ISOPHOT, ISOCAM and SWS on the $Infrared$ $Space$ $Telescope$ (ISO; e.g., \citealt{lut98}; \citealt{gen98}; \citealt{rig99}; \citealt{tra01}) and more recently with the infrared spectrograph (IRS; \citealt{hou04}) on the $Spitzer~ Space~ Telescope$ (\citealt{wee05}; \citealt{brl06}; \citealt{arm07}; \citealt{des07}). $Spitzer$ IRS has also extended these studies to fainter sources at higher redshifts. A particularly interesting population of high redshift optically faint infrared luminous sources (\citealt{yan04}; \citealt{bra07}; \citealt{fio08}; \citealt{dey08}) has been discovered using the Multiband Imaging Photometer (MIPS; \citealt{rie04}). Various observing programs with $Spitzer$ IRS have found that MIPS sources at flux density levels of f$_{\nu}$(24$\rm \mu m$) $\ge$ 0.8 mJy with optical magnitudes $R$ $\ga$ 24 are typically at z $\sim$ 2$-$3 (\citealt{hou05}; \citealt{wee06} \citealt{yan07}; \citealt{saj07}). The infrared spectra exhibit a wide range of properties from deep 9.7$\rm \mu m$ silicate absorption features and/or strong polycyclic aromatic hydrocarbon (PAH) emission features to featureless power-law continua. This obscured population is important in determining how much light originates from starbursts and AGN at epochs in the Universe when their luminosity density was at its maximum. Given that their optical faintness makes optical follow-up difficult, understanding how to characterize these sources using their infrared spectra is important in understanding their energetics.

One approach in understanding the contribution of starbursts and AGN to the infrared emission of optically faint infrared sources is to investigate the infrared spectra of sources also selected in other wavebands. For example, at one extreme we can study sources selected on the basis of large far-infrared flux densities, which we might expect to contain cool dust and show signatures of starbursts. In \citet{bra08}, we presented the IRS spectra of 11 70$\rm \mu m$-selected optically faint infrared luminous galaxies and found that seven had large 6.2$\rm \mu m$ PAH equivalent widths, consistent with them being powerful starburst galaxies. Four of the galaxies showed deep silicate absorption features and their properties suggested that AGN are likely to be the dominant origin of their large mid-infrared luminosities. 

At the other extreme, we can consider sources selected on the basis of X-ray flux, which we might expect to show spectral signatures of AGN. \citet{wee06b} presented IRS spectra of 9 X-ray-loud optically faint infrared luminous galaxies selected from the $Spitzer$ Wide-Area Infrared Extragalactic Survey (SWIRE; \citealt{lon04}). Eight sources have silicate absorption features and one source has a featureless power-law mid-IR spectrum.

In this paper, we present $Spitzer$ IRS spectra of 16 optically faint infrared galaxies in the Bo\"otes field of the NOAO Deep Wide-Field Survey (NDWFS; \citealt{jan99}) which are selected to be X-ray loud. This significantly increases the number of X-ray loud optically faint infrared luminous galaxies thus far studied, and extends the sample to higher X-ray and 24$\rm \mu m$ luminosities.  

A cosmology of $H_0 = 70 {\rm ~km ~s^{-1} ~Mpc^{-1}}$, $\Omega_M$=0.3, and $\Omega_\Lambda$=0.7 is assumed throughout. 

\section{Source Selection}

Our sample is selected from the Bo\"otes field of the NOAO Deep Wide-Field Survey (NDWFS; \citealt{jan99}). The combination of the large survey area (8.2 deg$^{2}$ with overlapping MIPS and optical coverage) and existing $Spitzer$ MIPS and IRAC, $Chandra$, and deep optical imaging data makes this an ideal data-set for multi-wavelength comparisons of sources.  The MIPS data were obtained with an effective integration time at 24$\rm \mu m$ of $\sim$90\,s per sky pixel, reaching a 5 $\sigma$ detection limit of $\sim$ 0.3 mJy for unresolved sources. The resulting source lists contain $\approx$28,000 sources with f$_{\nu}$(24$\rm \mu m$)$>$0.3 mJy. The XBo\"otes survey was undertaken using ACIS-I on the $Chandra X-ray Observatory$ with 5 ks extosures over the entire Bo\"otes field (\citealt{mur05}; \citealt{ken05}; \citealt{bra06}). It comprises 3293 point sources with 4 or more X-ray counts, corresponding to a limiting flux of f$_{(0.5-7 keV)}$=8.1$\rm \times 10^{-15}~ergs~cm^{-2}~s^{-1}$. An 8.5 deg$^2$ region of the Bo\"otes field has been mapped in all four $Spitzer$ IRAC bands by the IRAC Shallow Survey \citep{eis04}, reaching 5$\sigma$ limits of 6.4, 8.8, 51, and 50 $\rm \mu Jy$ at 3.6, 4.5, 5.8, and 8 $\rm \mu m$, respectively.

Using the XBo\"otes and MIPS 24$\rm \mu m$ catalogs, we selected a sample of 16 X-ray loud sources in the Bo\"otes field for $Spitzer$ IRS observations. We selected all 13 of the sources with $>$4 X-ray counts (f$_{0.5-7 keV} > 8.1 \times 10^{-15} \rm ~ergs~cm^{-2}~s^{-1}$), f$_{\nu}$(24$\rm \mu m$)$>$ 2 mJy, and $R>22$ mag (the ``main'' sample). We supplemented this sample with 3 extra sources (BootesX13, 14, and 15) which were slightly fainter at 24 $\rm \mu m$ (f$_{\nu}$(24$\rm \mu m$)$>$1 mJy) but which were particularly powerful X-ray sources (hereafter denoted as the ``X-ray bright'' sub-sample with 48, 34, and 19 X-ray counts corresponding to f$_{0.5-7 keV}$= 9.2, 6.5, and 3.7 $\times 10^{-14} \rm ergs~cm^{-2}~s^{-1}$ respectively). Because of their different selection criteria, we identify the X-ray bright sub-sample separately throughout the analysis and consider the conclusions with and without their inclusion.

The basic optical and infrared properties of the sample are shown in Table~\ref{tab:70sample}. 10/16 of the sources have fairly red optical to infrared colors and qualify as Dust Obscured Galaxies (DOGs) based in the criterion ($R$-[24]$>$14) presented in \citealt{dey08}. None of our sample have redshifts from optical spectroscopy. The X-ray properties are presented in Table~\ref{tab:xabs}. At $z>1$, the X-ray luminosities are far larger than can be attributed to a starburst and must be attributed to an AGN. At $z=0.9$ (the lowest measured source redshift; see below), the minimum X-ray flux, f$_{0.5-7 keV} = 8.1 \times 10^{-15} \rm ~ergs~cm^{-2}~s^{-1}$ corresponds to an X-ray luminosity of $\approx 3 \times 10^{43}$ ergs s$^{-1}$. If this was all due to star-formation, it would imply an unfeasibly large star-formation rate of $\approx$ 5000 M$\odot$ yr$^{-1}$ \citep{gri03}. 
\begin{deluxetable*}{llllllllllllll}
\tabletypesize{\scriptsize}
\setlength{\tabcolsep}{0.05in}
\tablecolumns{14} 
\tablewidth{0pc}
\tablecaption{\label{tab:70sample} Optical and IR properties.} 
\tablehead{ 
\colhead{IRS ID} & \colhead{MIPS} & \colhead{$B_W$$^{a,b}$} & \colhead{$R$$^{a,b}$} & \colhead{$I$$^{a,b}$} & \colhead{$K$$^{a,b}$} & \colhead{f$_\nu$3.6$^c$} & \colhead{f$_\nu$4.5$^c$} & \colhead{f$_\nu$5.8$^c$} & \colhead{f$_\nu$8$^c$} & \colhead{f$_{\nu}$24$^c$} & \colhead{f$_\nu$70} & \colhead{f$_\nu$160} &\colhead{R-[24]}\\
\colhead{} & \colhead{name} & \colhead{mag} & \colhead{mag} & \colhead{mag} & \colhead{mag} & \colhead{mJy} & \colhead{mJy} & \colhead{mJy} & \colhead{mJy} & \colhead{mJy} & \colhead{mJy} & \colhead{mJy} & \colhead{mag}\\
}
\startdata
BootesX1    &SST24J143048.3+322532  & 24.00   &23.11  &22.33  &$-$     & 0.13 &  0.22 & 0.77 &   0.68 & 3.15 & $<$12 &$<$120& 14.70 \\
BootesX2    &SST24J142731.3+324434  & 25.54   &23.04  &21.51  &$-$     & 0.09 &  0.12 & 0.20 &   0.59 & 2.61 & $<$15 &$<$100& 14.42\\ 
BootesX3    &SST24J142609.0+333303  & 23.33   &22.80  &21.84  &18.29$^d$& 0.05 & 0.06 & 0.10 &   0.18 & 2.49 & $<$20 &$<$120& 14.13\\
BootesX4    &SST24J143234.9+333637   &24.53   &22.96  &21.53  &18.65$^d$& 0.18 & 0.35 & 0.73 &   1.68 & 2.92 &30$\pm$6&110$\pm$22& 14.47\\
BootesX5    &SST24J142625.5+334051   &25.14   &23.53  &22.04  &18.00$^d$& 0.21 & 0.32 & 0.52 &   0.82 & 2.11 & $-$   &$<$120& 14.69\\
BootesX6    &SST24J143232.6+335903   &23.09   &22.79  &22.10  &17.85$^d$& 0.04 & 0.05 & 0.16 &   0.38 & 2.42 &10$\pm$2&80$\pm$16& 14.09\\ 
BootesX7    &SST24J143038.4+340929   &24.34   &22.86  &21.67  &18.53$^d$& 0.06 & 0.10 & 0.17 &   0.35 & 2.40 &12$\pm$2&$<$100& 14.15\\
BootesX8    &SST24J143301.5+342341   &23.48   &22.85  &22.54  &18.32$^d$& 0.10 & 0.16 & 0.35 &   0.62 & 2.17 & $<$15 &$<$120& 14.04\\
BootesX9    &SST24J142849.0+343432   &23.58   &22.35  &21.14  &18.11$^e$& 0.08 & 0.10 & 0.13 &   0.33 & 2.36 & $<$15 &$<$100& 13.60\\
BootesX10   &SST24J143834.9+343839   &22.76   &22.44  &22.58  & $-$     & 0.06 &-1.0$^g$&0.25&-1.0$^g$& 2.25 & $<$20 &$<$150& 13.66\\
BootesX11   &SST24J143715.1+345323   &24.15   &23.73  &23.06  & $-$     & 0.16 & 0.27 & 0.49 &   0.95 & 2.51 & $<$15 &$<$120& 15.07\\
BootesX12   &SST24J142537.8+351735   &24.21   &22.12  &20.69  &18.30$^e$& 0.09 & 0.12 & 0.21 &   0.38 & 2.28 &25$\pm$5&$<$120& 13.36\\
Houck13     &SST24J143644.2+350627   &24.71   &23.69  &23.36  &19.11$^f$& 0.04 & 0.10 & 0.32 &   0.69 & 2.34 &9$\pm$3$^h$&43$\pm$12$^h$& 14.96\\
\cutinhead{X-ray bright sub-sample}
BootesX13   &SST24J143723.4+334308   &24.96   &22.78  &22.49  & $-$     & 0.03 & 0.03 & 0.04 &   0.11 & 1.44 &20$\pm$4&$<$120& 13.52\\
BootesX14   &SST24J142707.0+325213   &$-$     &22.03  &$-$    & $-$     & 0.19 & 0.16 & 0.19 &   0.25 & 1.22 & $<$12 &$<$100&12.59\\
BootesX15   &SST24J143359.0+331301   &23.95   &22.40  &21.06  &17.15$^d$& 0.23 & 0.29 & 0.42 &   0.61 & 1.92 & $<$15 &$<$120& 13.45\\
\enddata
\tablenotetext{a}{All quoted magnitudes are Vega magnitudes from the NDWFS DR3 (Jannuzi et al. in preparation).}
\tablenotetext{b}{Errors on $B_w$, $R$, $I$, and $K$-band magnitudes are $<$0.1.}
\tablenotetext{c}{Errors on f$_\nu$(3.6), f$_\nu$(4.5), f$_\nu$(5.8), f$_\nu$(8), and f$_\nu$(24) flux densities are $<$0.1 mJy.}
\tablenotetext{d}{$K_s$-mag from FLAMEX survey \citep{els06}.}
\tablenotetext{e}{$K$-mag from NDWFS survey (Dey et al. in preparation.)}
\tablenotetext{f}{$K$-mag from Keck/NIRC observation.}
\tablenotetext{g}{Because BootesX10 lies at the edge of our IRAC coverage, no detections were found in this band.}
\tablenotetext{h}{Value from deeper imaging data presented in Tyler et al. submitted.}
\end{deluxetable*} 

 \begin{deluxetable*}{lllllll}
\tabletypesize{\scriptsize}
\setlength{\tabcolsep}{0.1in}
\tablecolumns{7} 
\tablewidth{0pc}
\tablecaption{\label{tab:xabs} X-ray properties.} 
\tablehead{ 
\colhead{IRS ID} & \colhead{X-ray} & \colhead{X-ray$^a$} & \colhead{hardness} & \colhead{fx$^c$}  & \colhead{log$_{10}$(fx / f$_{opt}$)$^d$} & \colhead{log(Lx$^e$)}\\
\colhead{} & \colhead{name} & \colhead{counts} & \colhead{ratio$^b$} & \colhead{} & \colhead{} & \colhead{($\rm erg~s^{-1}$)}  \\
}
\startdata  
BootesX1          & CXOXB J143048.3+322531   &   5$\pm$2.2   &   1.00$\pm_{0.48}^{0.00}$  & 8.5$\pm$5.6    &   0.67 &   44.22\\
BootesX2          & CXOXB J142731.4+324435   &   6$\pm$2.4   &    0.68$\pm_{0.25}^{0.19}$ & 10.6$\pm$5.9   &  0.74  &   43.63\\
BootesX3          & CXOXB J142609.0+333304   &   5$\pm$2.2   &    0.20$\pm_{0.28}^{0.26}$ &  9.5$\pm$5.9   &  0.60  &   44.44\\
BootesX4          &CXOXB J143234.9+333636    &   4$\pm$2.0   &    0.51$\pm_{0.37}^{0.29}$ &  8.1$\pm$5.6   &   0.59 &   43.75\\
BootesX5          &CXOXB J142625.5+334052    &   7$\pm$2.6   &$-$0.16$\pm_{0.19}^{0.20}$  & 14.3$\pm$6.7   &   1.07 &   44.23  \\
BootesX6          &CXOXB J143232.6+335902    &  21$\pm$4.6   &$-$0.44$\pm_{0.06}^{0.07}$  & 43.8$\pm$10.0  &   1.26 &  44.25\\
BootesX7          &CXOXB J143038.5+340928    &  7$\pm$2.6    &    0.72$\pm_{0.22}^{0.16}$ & 14.0$\pm$6.7   &   0.79 &   43.97\\
BootesX8          &CXOXB J143301.5+342343    & 14$\pm$3.7   &    0.00$\pm_{0.10}^{0.10}$  & 27.1$\pm$8.4   &  1.07  &   44.95 \\
BootesX9          &CXOXB J142849.0+343432    &  6$\pm$2.4    &    0.67$\pm_{0.25}^{0.19}$ & 12.5$\pm$6.5   &   0.54 &      $-$\\
BootesX10         &CXOXB J143835.0+343839    &  9$\pm$3.0    &    0.11$\pm_{0.15}^{0.15}$ & 18.0$\pm$7.3   &   0.73 &   44.99\\
BootesX11         &CXOXB J143715.1+345323    &  5$\pm$2.2    &$-$0.20$\pm_{0.26}^{0.27}$  & 20.5$\pm$6.0   &  1.30  &      $-$\\
BootesX12         &CXOXB J142537.8+351735    & 29$\pm$5.4   &$-$0.10$\pm_{0.05}^{ 0.05}$  &56.8$\pm$11.1   &  1.10 &      $-$\\
Houck13           &CXOXB J143644.2+350626    & 7$\pm$2.6    &     0.43$\pm_{0.20}^{0.18}$ &13.2$\pm$6.5    &  1.10 &   44.55\\
\cutinhead{X-ray bright sub-sample}
BootesX13         &CXOXB J143723.5+334307    & 48$\pm$6.9   &$-$0.21$\pm_{0.03}^{0.03}$   & 92.0$\pm$13.9  &  1.58 &      $-$\\
BootesX14         &CXOXB J142707.0+325212    & 34$\pm$5.8   &$-$0.47$\pm_{0.04}^{0.04}$   & 64.9$\pm$12.1  &  1.21 &      $-$\\
BootesX15         &CXOXB J143359.1+331300    &19$\pm$4.3    &    0.47$\pm_{0.07}^{0.07}$  & 37.2$\pm$9.6   &  1.03 &      $-$\\
\enddata
\tablenotetext{a}{Counts in observed total (0.5-7 keV) band and their associated Poisson errors.}
\tablenotetext{b}{The Hardness ratio is defined as H-S/H+S where H denotes the X-ray counts in the hard (2-7 keV) band and S denotes the X-ray counts in the soft (0.5-2 keV) band. }
\tablenotetext{c}{Observed 0.5-7 keV X-ray flux and error in 10$^{-15}\rm ~ergs~cm^{-2}~s^{-1}$ assuming a power-law spectrum with photon index $\Gamma$=1.7 and a galactic column density of $1\times10^{20}\rm~cm^{2}$ \citep{sta92}. No correction has been made for intrinsic X-ray absorption.} 
\tablenotetext{d}{X-ray to optical ($R$-band) flux ratio defined as log$_{10}$(fx / f$_{opt}$)=log$_{10}$(fx) + 0.4$R$ +5.5 \citep{bra06}}
\tablenotetext{e}{Rest-frame 0.5-7 keV X-ray luminosity calculated using redshifts in Table~3.}

\end{deluxetable*} 

\section{Observations and Data Analysis}

The spectroscopic observations were made with the IRS Short Low module in order 1 only (SL1) and with the Long Low module in orders 1 and 2 (LL1 and LL2), described in \citet{hou04}.  These orders give low resolution spectral coverage from $\sim$8\,$\mu$m to $\sim$35\,$\mu$m. The integration times for individual sources are given in Table~\ref{tab:irs}. Data were processed with version 13.0 of the SSC pipeline, and extraction of the source spectra was done with the SMART analysis package \citep{hig04}. For details of the observational and data analysis techniques, we refer the reader to \citet{bra08}. 

\section{Results}

\begin{figure*}[h]
\begin{center}
\includegraphics[height=48mm]{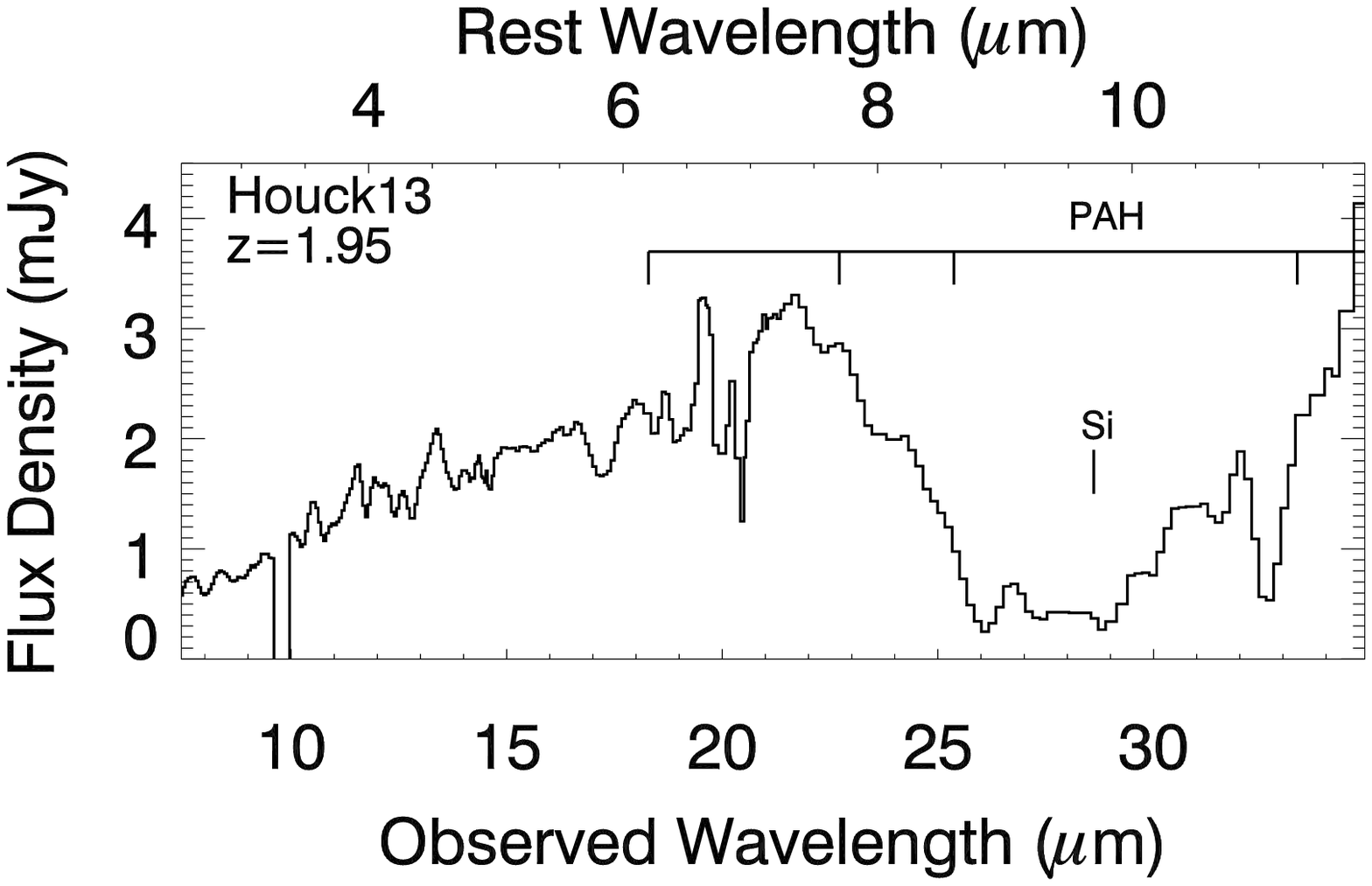}
\includegraphics[height=48mm]{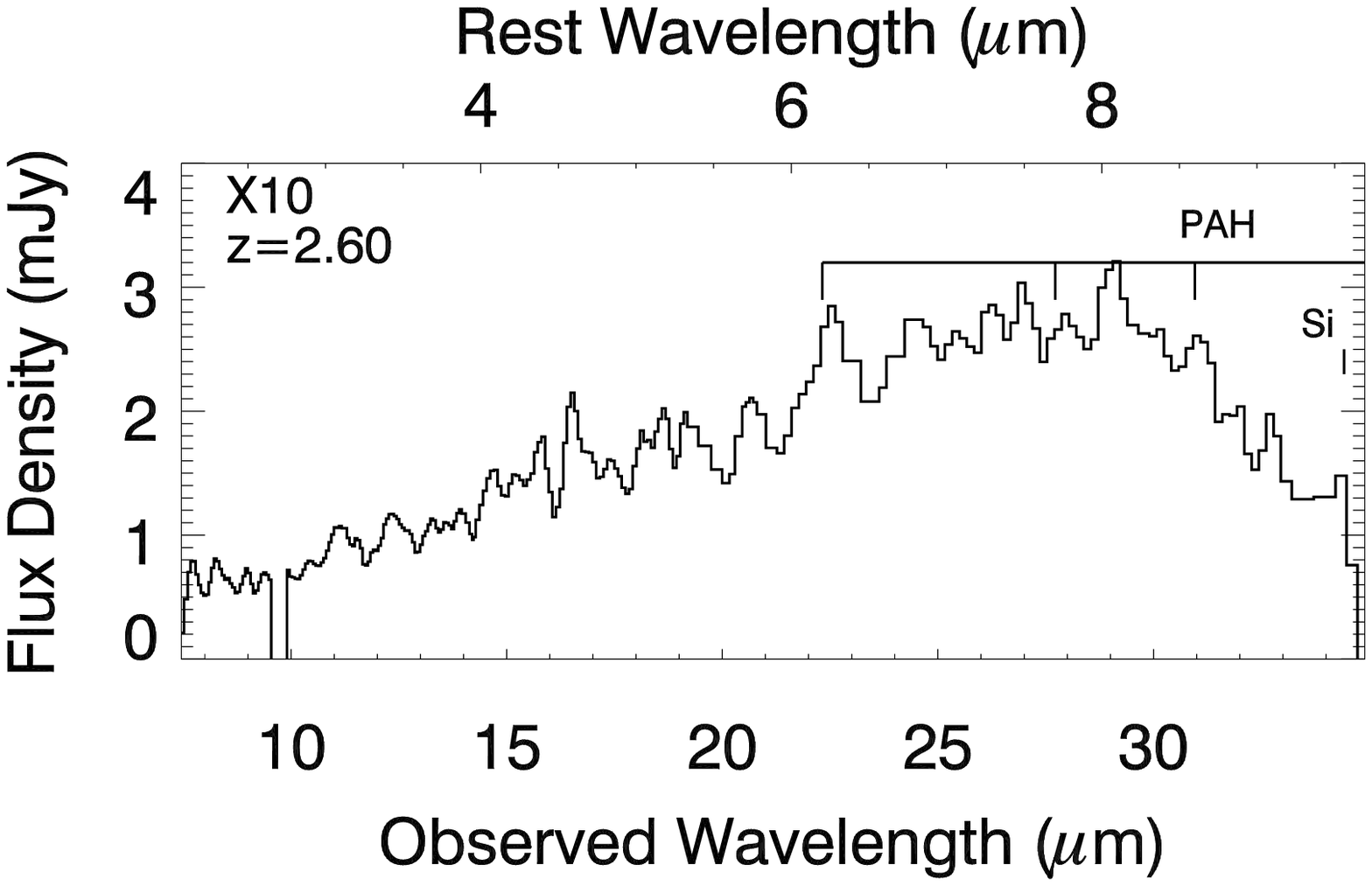}
\includegraphics[height=48mm]{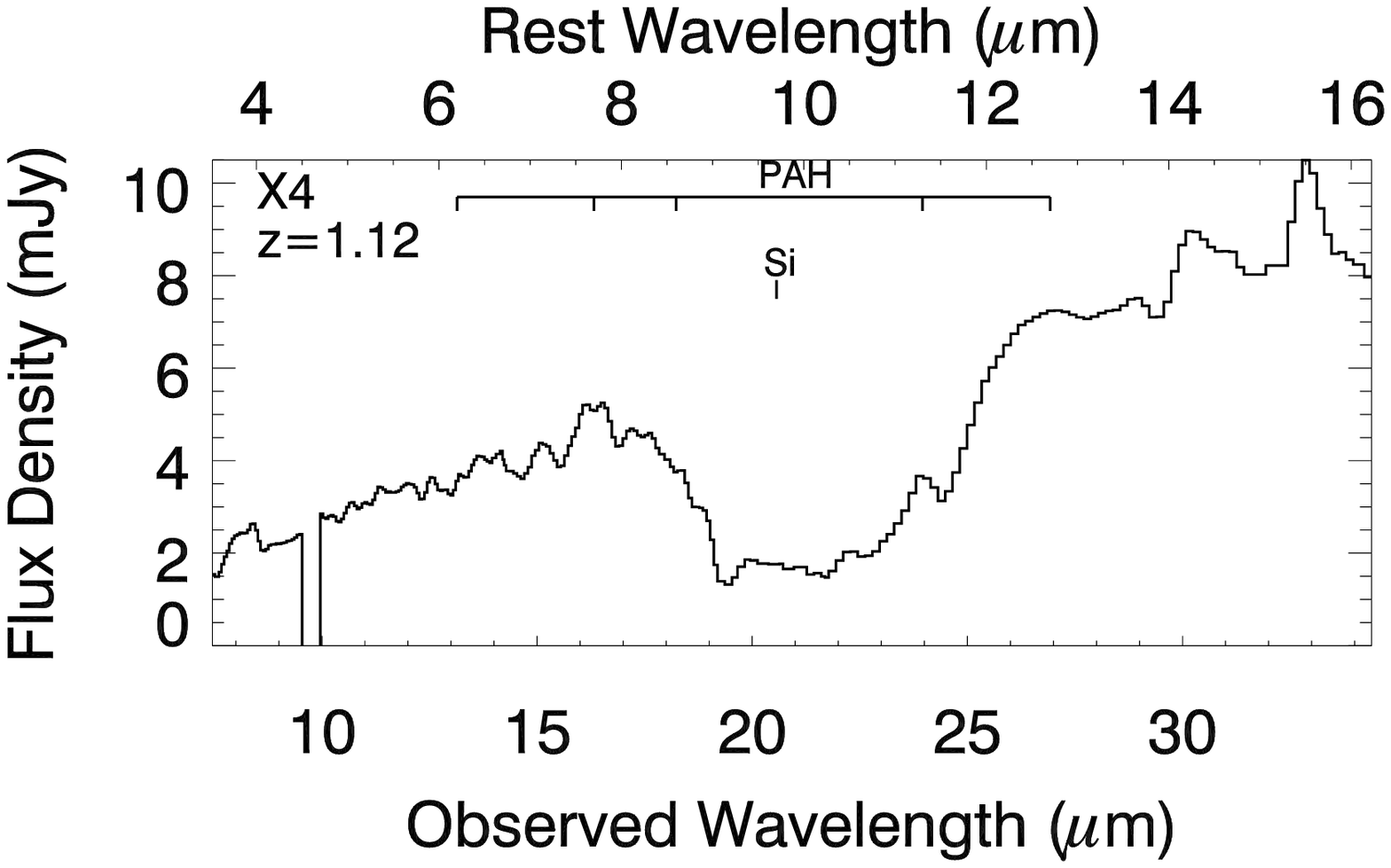}
\includegraphics[height=48mm]{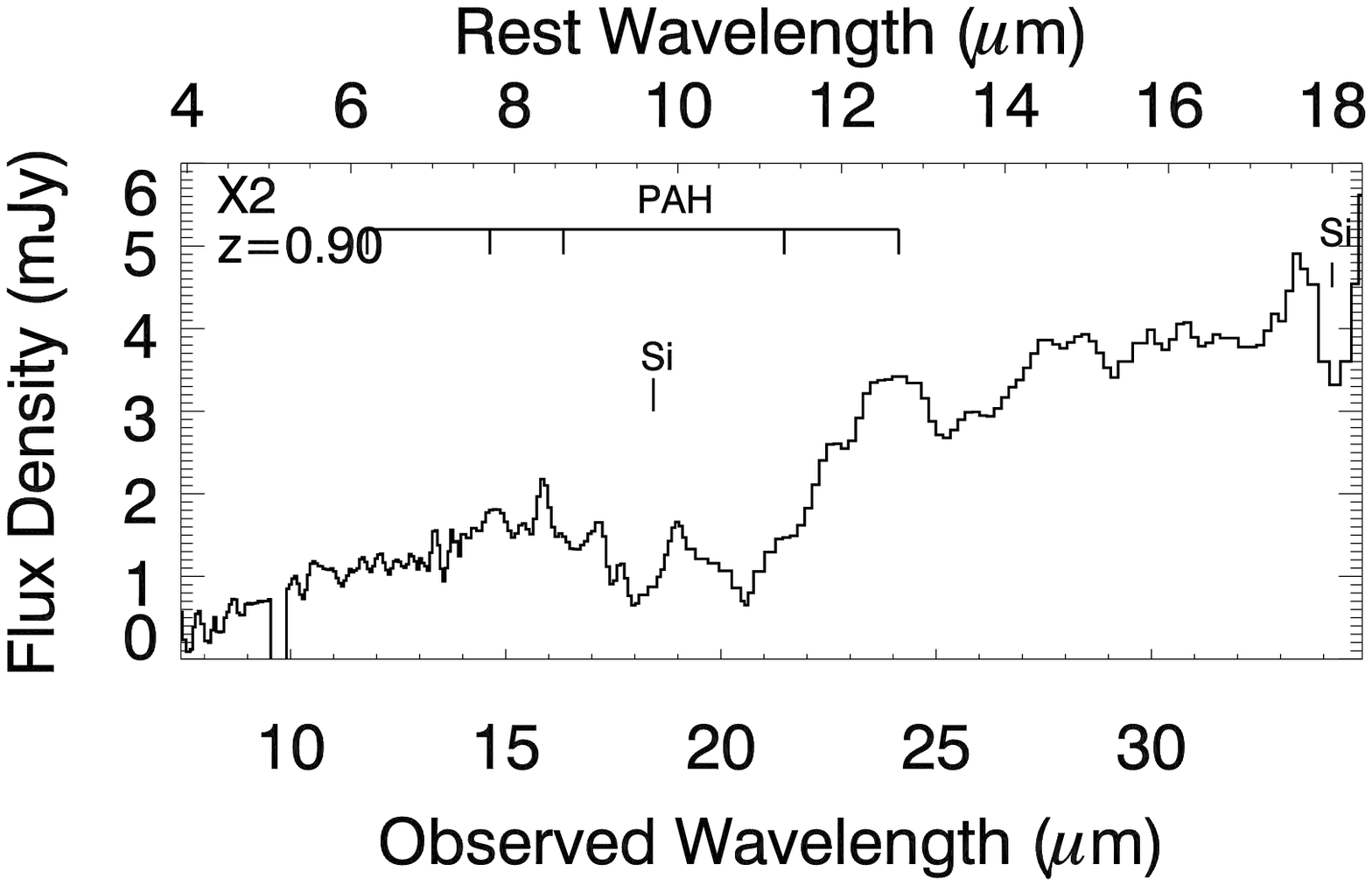}
\includegraphics[height=48mm]{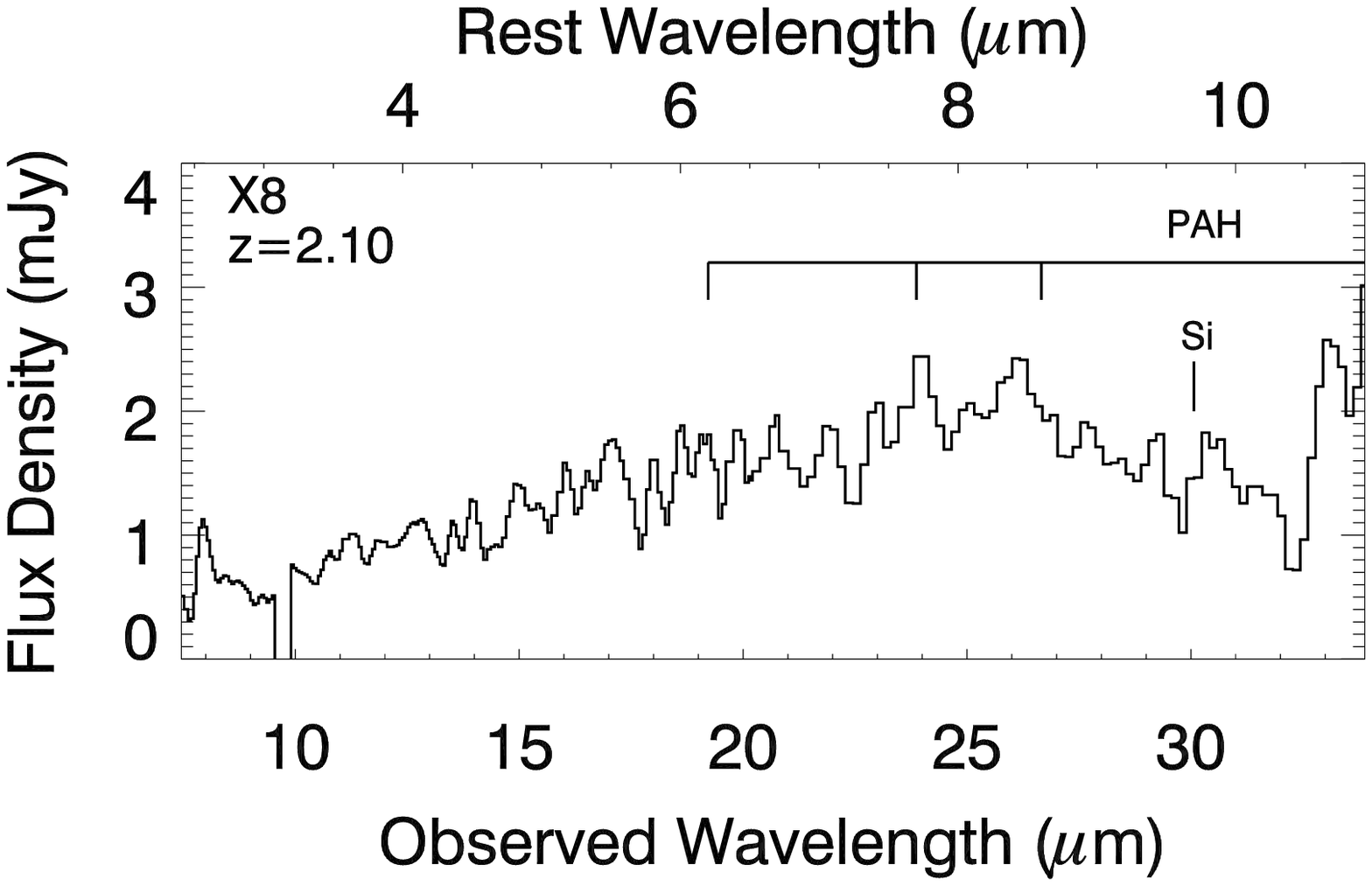}
\includegraphics[height=48mm]{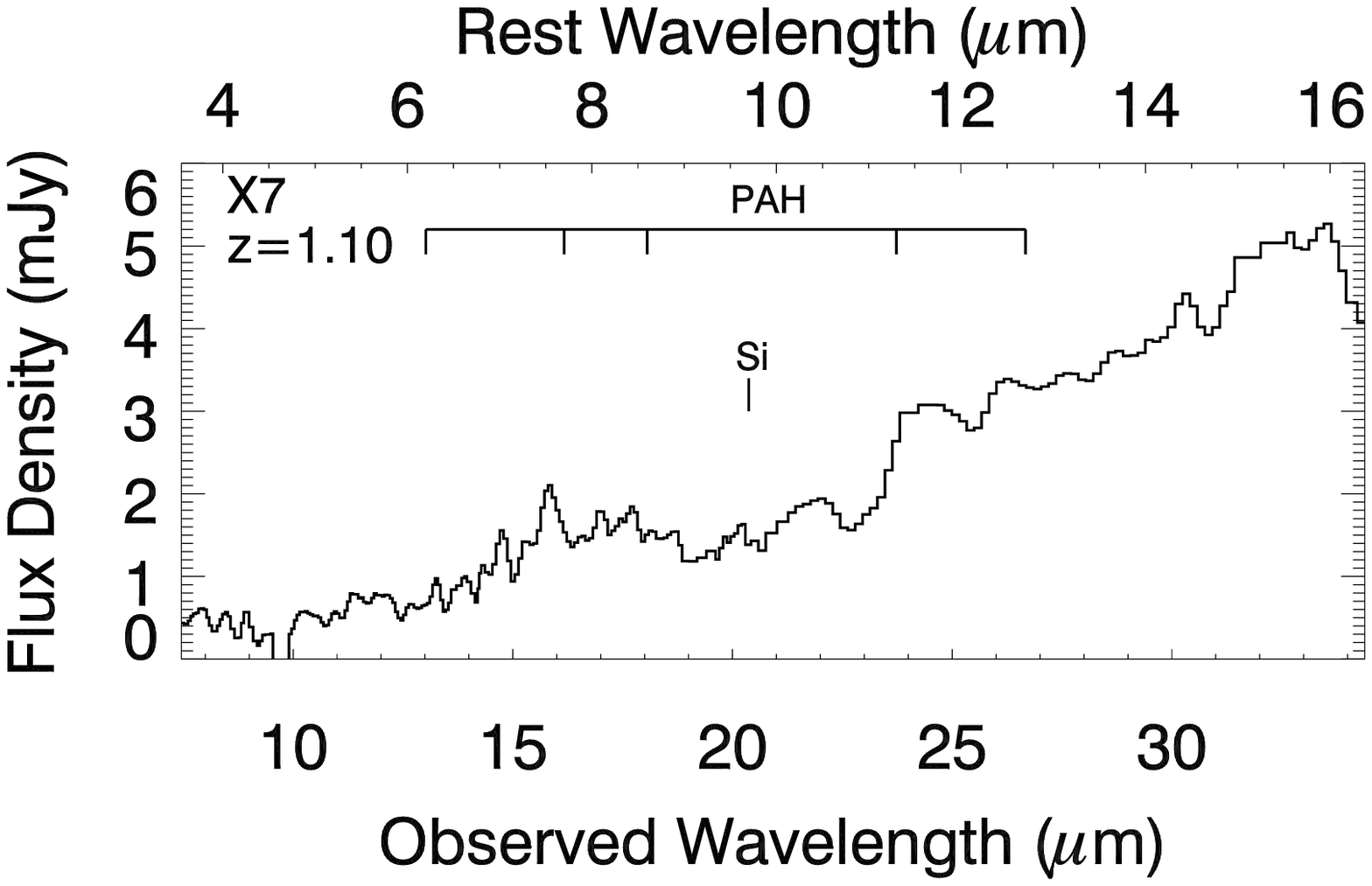}
\includegraphics[height=48mm]{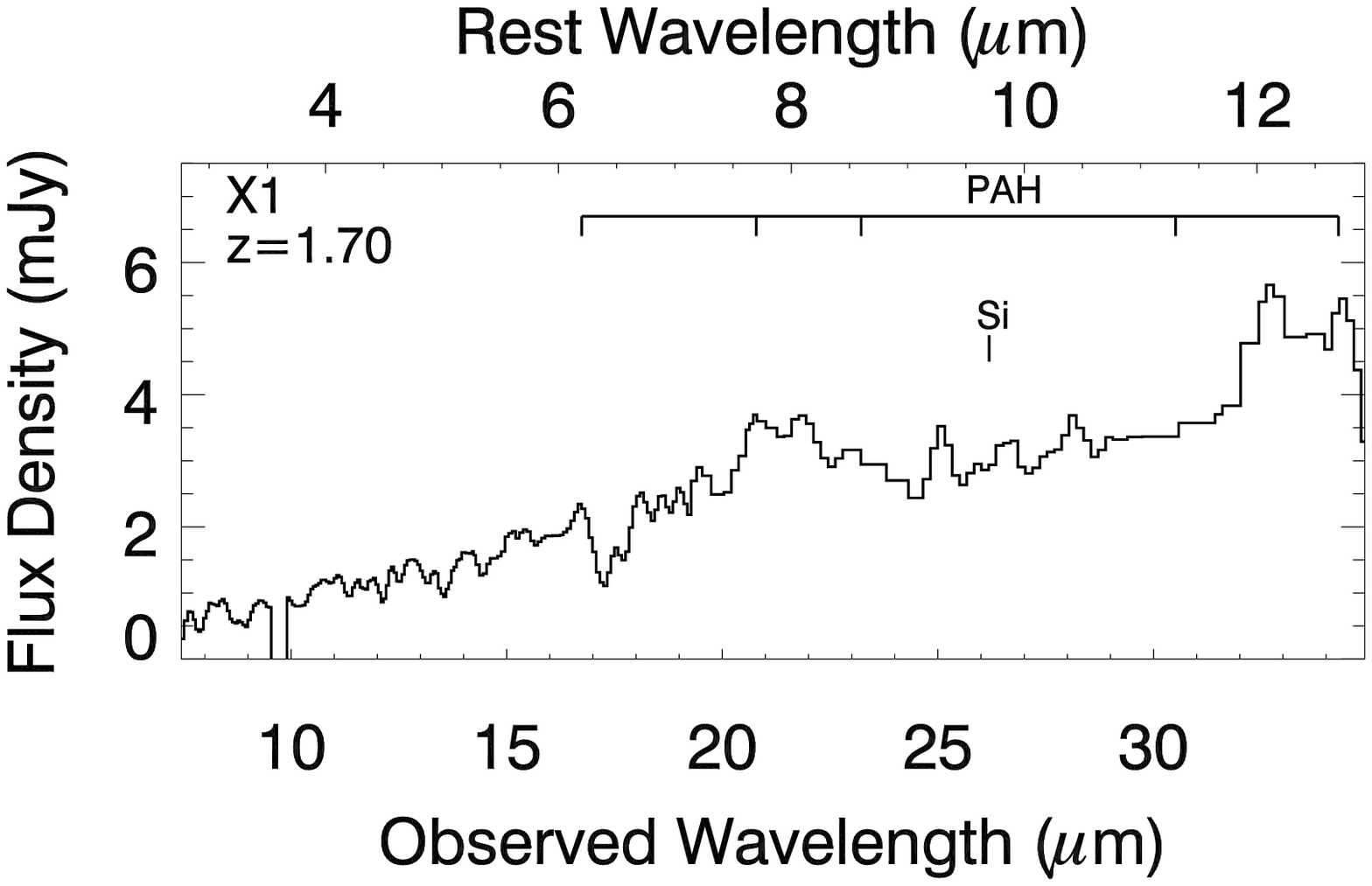}
\includegraphics[height=48mm]{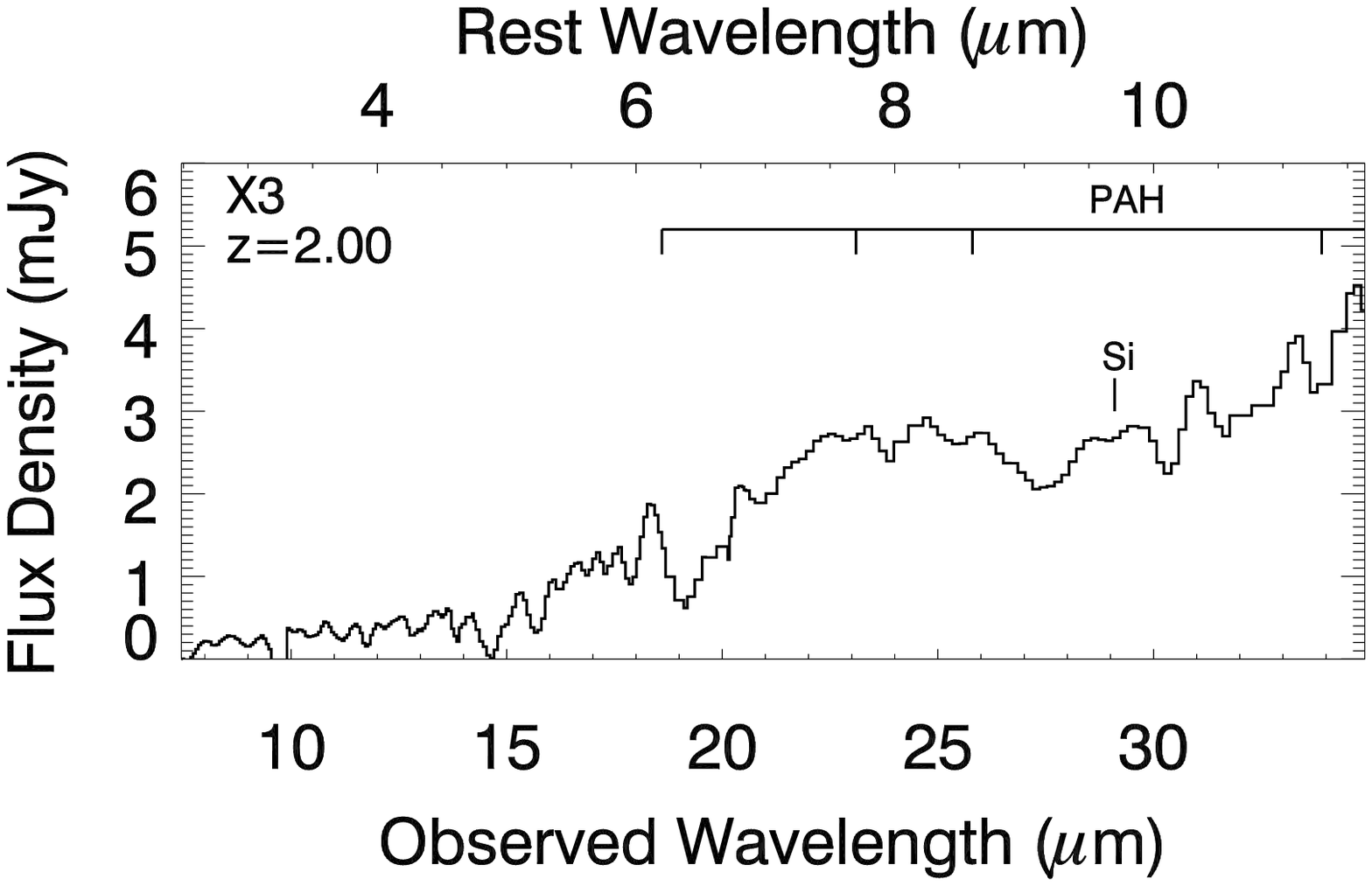}
\end{center}
{\caption[junk]{\label{fig:xspec} IRS spectra of X-ray selected optically faint infrared luminous galaxies ordered by their silicate absorption strengths. The IRS spectrum of Houck13 was published in \citet{hou05}. The spectra are boxcar smoothed over a resolution element (approximately two pixels). In cases where we have an estimated redshift, the expected positions of the PAH emission features and silicate absorption features are shown. A typical noise spectrum of IRS observations with similar 24 $\rm \mu m$ flux densities and exposure times can be found in \citet{bra08}.}}
\end{figure*}

The IRS spectra of all 16 X-ray selected sources are shown in Figure~\ref{fig:xspec}. Although the IRS spectrum of Houck13 was presented in \citet{hou05}, it satisfies the selection criteria of the main sample and we include it again here for completeness. 9/13 of the main sample (and 9/16 of the total ``main + X-ray bright'' sample) exhibit identifiable 9.7$\rm \mu m$ silicate absorption features (S$_{10}<$-0.4\footnote{S$_{10}$ is defined in Table~\ref{tab:irs}}; hereafter described as ``absorption-dominated'' sources and denoted by ``abs'' in Table~\ref{tab:irs}). A further 4/13 or the main sample (and 7/16 of the total ``main + X-ray bright'' sample) exhibit power-law spectra with no convincing features (hereafter described as ``power-law'' sources and denoted by ``pow'' in Table~\ref{tab:irs}); either they have only weak silicate absorption features or they are at $z>2.6$ and the absorption feature does not fall within the observable IRS bandwidth. BootesX6 may have a weak silicate emission feature if it is at $z$=0.9. Given that this is highly uncertain without other convincing features in the spectrum, we classify BootesX6 as a power-law source. None of the sources show convincing PAH emission features, although, given their uncertain or unknown redshifts, there may be weak PAH emission in some individual sources. We discuss each class in turn.

\begin{deluxetable*}{lllllllllllll}
\tabletypesize{\scriptsize}
\setlength{\tabcolsep}{0.05in}
\tablecolumns{13}
\tablewidth{0pc}
\tablecaption{\label{tab:irs} IRS and derived physical properties.} 
\tablehead{ 
\colhead{IRS ID} & \colhead{e.time} & \colhead{e.time} & \colhead{IRS} & \colhead{$\rm\nu L_{\nu}6\mu m$}  & \colhead{$\rm\nu L_{\nu}8\mu m$} & \colhead{L$_{IR}$$^c$} & \colhead{f$_\nu (15)$} & \colhead{S$_{10}$$^b$} & \colhead{PAH EW$^e$} & \colhead{SFR$^{g}$} & \colhead{L$_{IR}$(SFR)$^{h}$} & \colhead{IRS} \\
\colhead{} & \colhead{SL$^a$ (s)} & \colhead{LL$^f$ (s)} & \colhead{$z$} & \colhead{log(L$_\odot$)} & \colhead{log(L$_\odot$)} & \colhead{log(L$_\odot$)} & \colhead{/f$_\nu (6)$} & \colhead{} & \colhead{($\rm \mu m$)} & \colhead{(M$_\odot$ yr$^{-1}$)} & \colhead{log(L$_\odot$)} & \colhead{class$^d$}  \\
}
\startdata
BootesX1    &240&480  &1.7    &  12.2   & 12.6 & 13.3 & $-$ & $-$0.4 & $<$0.05 & $<$1380 & $<$12.8 & abs \\
BootesX2    &360&720  &0.9    &  11.5   & 11.8 & 12.5 & 3.3 & $-$1.0 & $<$0.04 & $<$190  & $<$12.0 & abs\\
BootesX3    &480&960  & 2.0$^f$& 12.2   & 12.6 & 13.3 & $-$ & $-$0.4 & $<$0.20 & $<$1170 & $<$12.9 & abs\\
BootesX4    &360&720  & 1.12  &  12.2   & 12.4 & 13.1 & 2.4 & $-$1.4 & $<$0.04 & $<$450  & $<$12.4 & abs\\
BootesX5    &480&960  & 1.4   &  12.0   & 12.3 & 13.0 & 2.8 & $-$0.4 & $<$0.09 & $<$590  & $<$12.5 & abs\\
BootesX6    &360&720  & 0.9   &   $-$   & $-$  & $-$  & $-$ & $-$    & $-$     & $-$     & $-$     & pow/em?\\
BootesX7    &480&960  & 1.1   &  11.5   & 11.9 & 12.6 & 6.3 & $-$0.5 & $<$0.05 & $<$110  & $<$11.7 & abs \\
BootesX8    &480&960  & 2.1   &  12.3   & 12.5 & 13.2 & $-$ & $-$0.6 & $<$0.05 & $<$560  & $<$12.5 & abs\\
BootesX9    &480&960  & $-$   &   $-$   & $-$  & $-$  & $-$ & $-$    & $-$     & $-$     & $-$     & pow\\
BootesX10   &480&960  & 2.6   &  12.6   & 12.9 & 13.6 & $-$ &$<-1.8$ &$<$0.10  & $<$1670 & $<$12.9 & abs\\
BootesX11   &480&960  & $-$   &   $-$   & $-$  & $-$  & $-$ & $-$    & $-$     & $-$     & $-$     & pow\\
BootesX12   &480&960  & $-$   &   $-$   & $-$  & $-$  & $-$ & $-$    & $-$     & $-$     & $-$     & pow\\
Houck13     &480&720  & 1.95  &  12.3   & 12.5 & 13.2 & $-$ & $-$2.9 & $<$0.03 & $<$420  & $<$12.3 & abs\\
\cutinhead{X-ray bright sub-sample}
BootesX13   &480&1200 & $-$   &    $-$ & $-$  & $-$ & $-$  & $-$     & $-$ & $-$ & $-$ & pow\\
BootesX14   &480&1200 & $-$   &    $-$ & $-$  & $-$ & $-$  & $-$     & $-$ & $-$ & $-$ & pow\\
BootesX15   &480&1200 & $-$   &    $-$ & $-$  & $-$ & $-$  & $-$     & $-$ & $-$ & $-$ & pow\\
\enddata
\tablenotetext{a}{Total integration times in SL1.}
\tablenotetext{f}{Total integration times in each of LL1 and LL2.}
\tablenotetext{b}{Silicate absorption strength is defined as $S_{10}=ln\frac{f_{obs}(10 \rm \mu m)}{f_{cont}(10 \rm \mu m)}$, where $f_{obs}(10 \rm \mu m)$ is the observed flux density at the peak of the 10$\rm \mu m$ feature and $f_{cont}(10 \rm \mu m)$ is the continuum flux at the peak wavelength, extrapolated from the continuum to either side.}
\tablenotetext{c}{The total infrared luminosity, L$_{IR}$=L(8-1000$\rm \mu m$), as estimated from the measured rest-frame 8$\rm \mu m$ flux density using the relation for the template of Mrk 231 (L$_{IR}$=5.0$\times \nu L_{\nu}(8\mu m)$).}
\tablenotetext{d}{Absorption-dominated and featureless power-law sources are denoted "abs" and "pow" respectively. "em" denotes possible silicate emission.}
\tablenotetext{e}{Rest-frame equivalent width of 6.2$\rm \mu m$ PAH emission feature.}
\tablenotetext{f}{The redshift could be $z$=1.5 if instead of silicate absorption at an observed wavelength of 29\AA, the spectrum is interpreted as having silicate emission at an observed wavelength of $\approx$24\AA. However, silicate emission is generally only observed in type I AGN which tend to have much brighter optical magnitudes for such bright 24$\rm \mu m$ emission (e.g., \citealt{bro06}).}
\tablenotetext{g}{Maximum star formation rate (SFR) estimated from the measured limit on the 6.2$\rm \mu m$ PAH emission feature using the relation of \citet{pop07}.}
\tablenotetext{h}{Maximum infrared luminosity associated with star formation estimated from the measured limit on the 6.2$\rm \mu m$ PAH emission feature using the relation of \citet{pop07}.}
\end{deluxetable*} 

\begin{figure*}[h]
\begin{center}
\includegraphics[height=48mm]{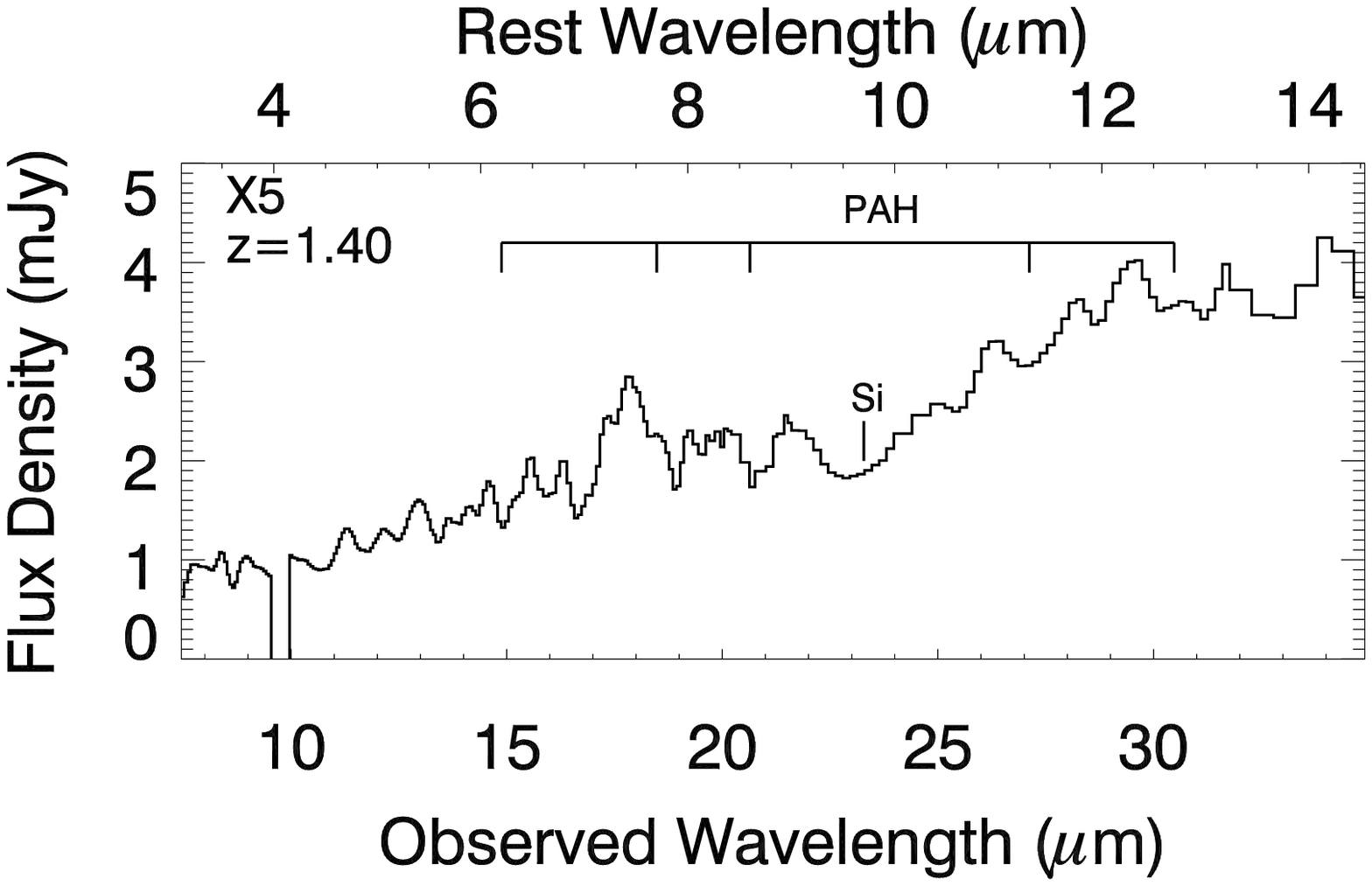}
\includegraphics[height=48mm]{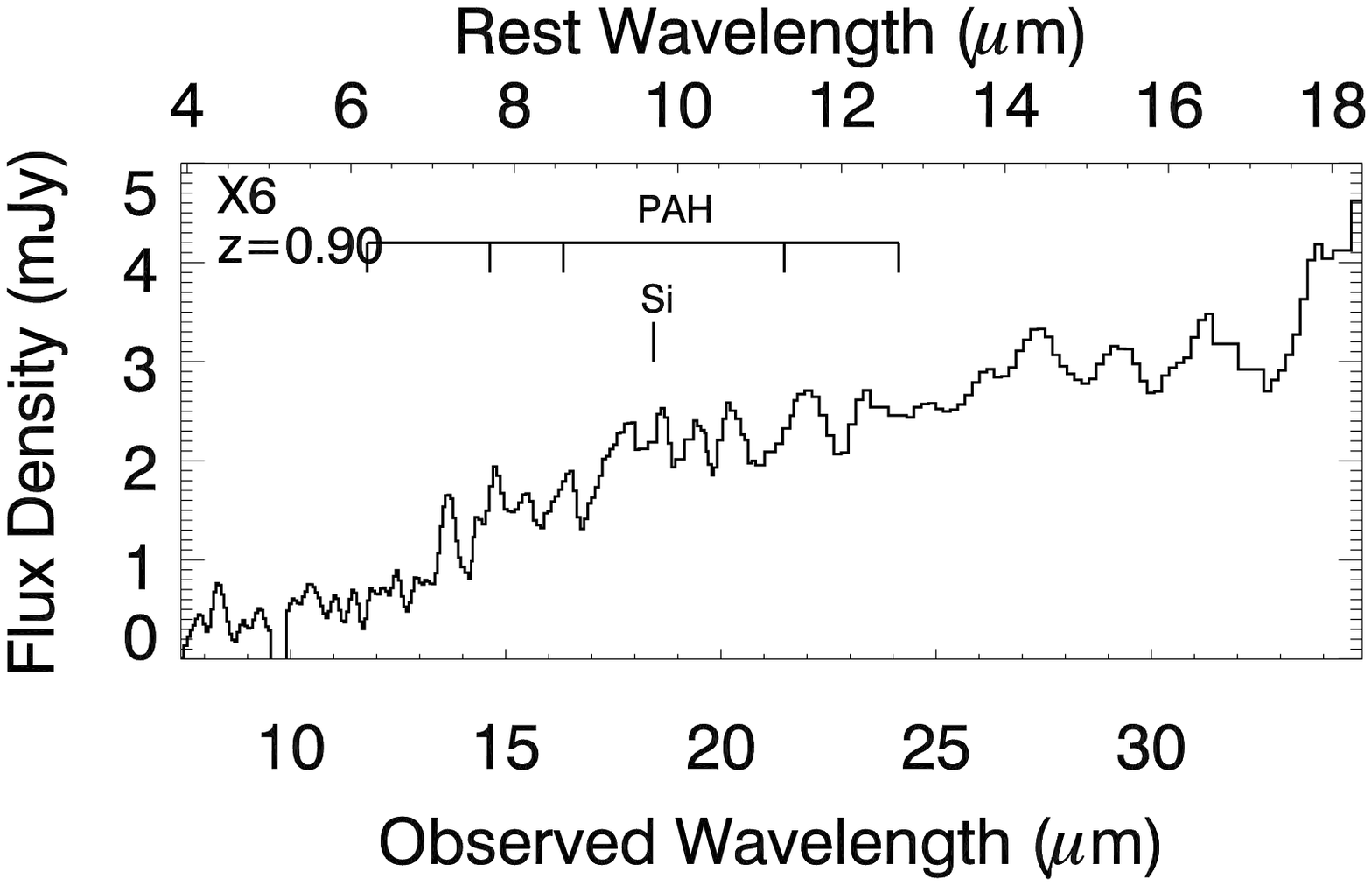}
\includegraphics[height=48mm]{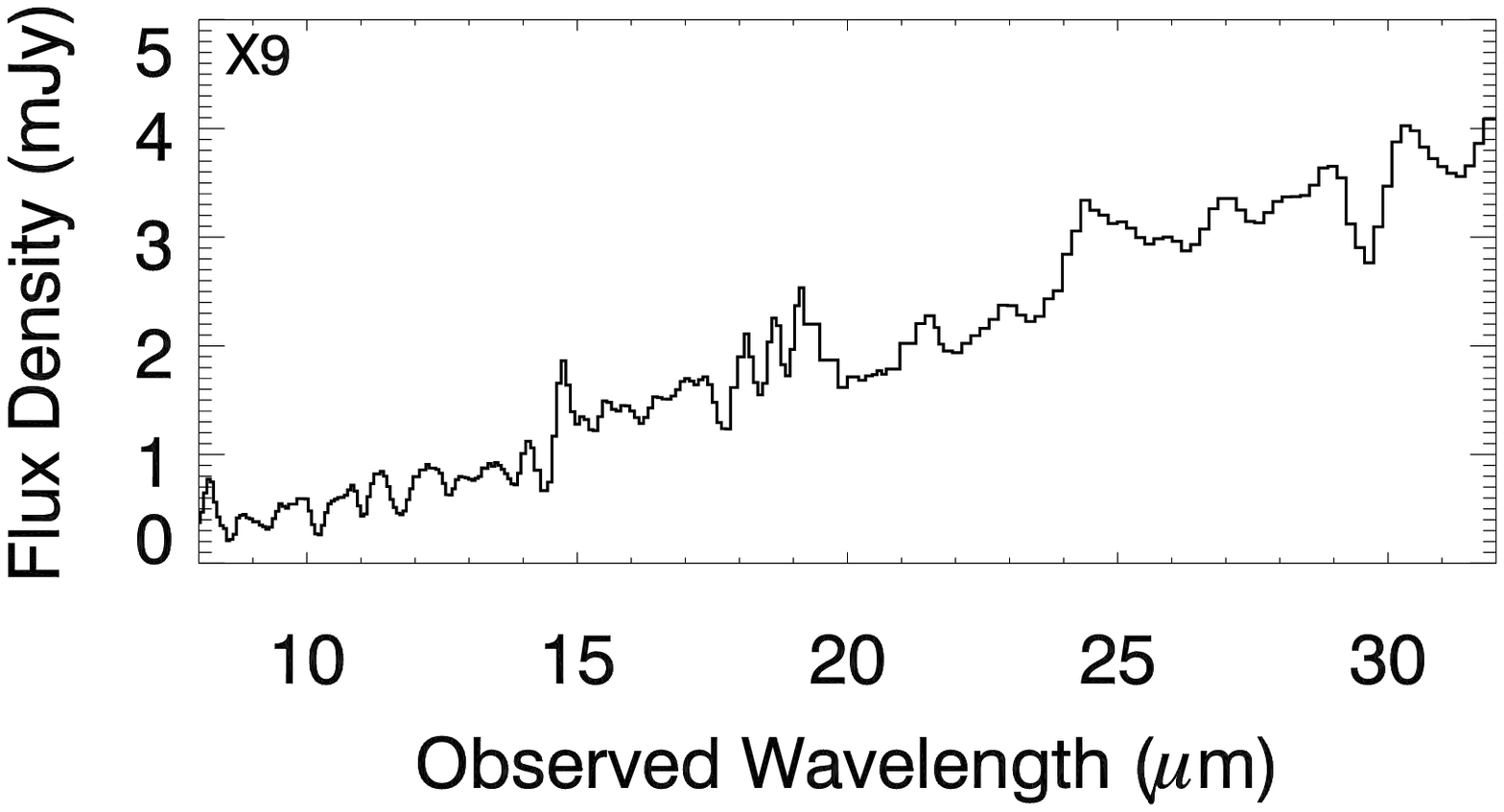}
\includegraphics[height=48mm]{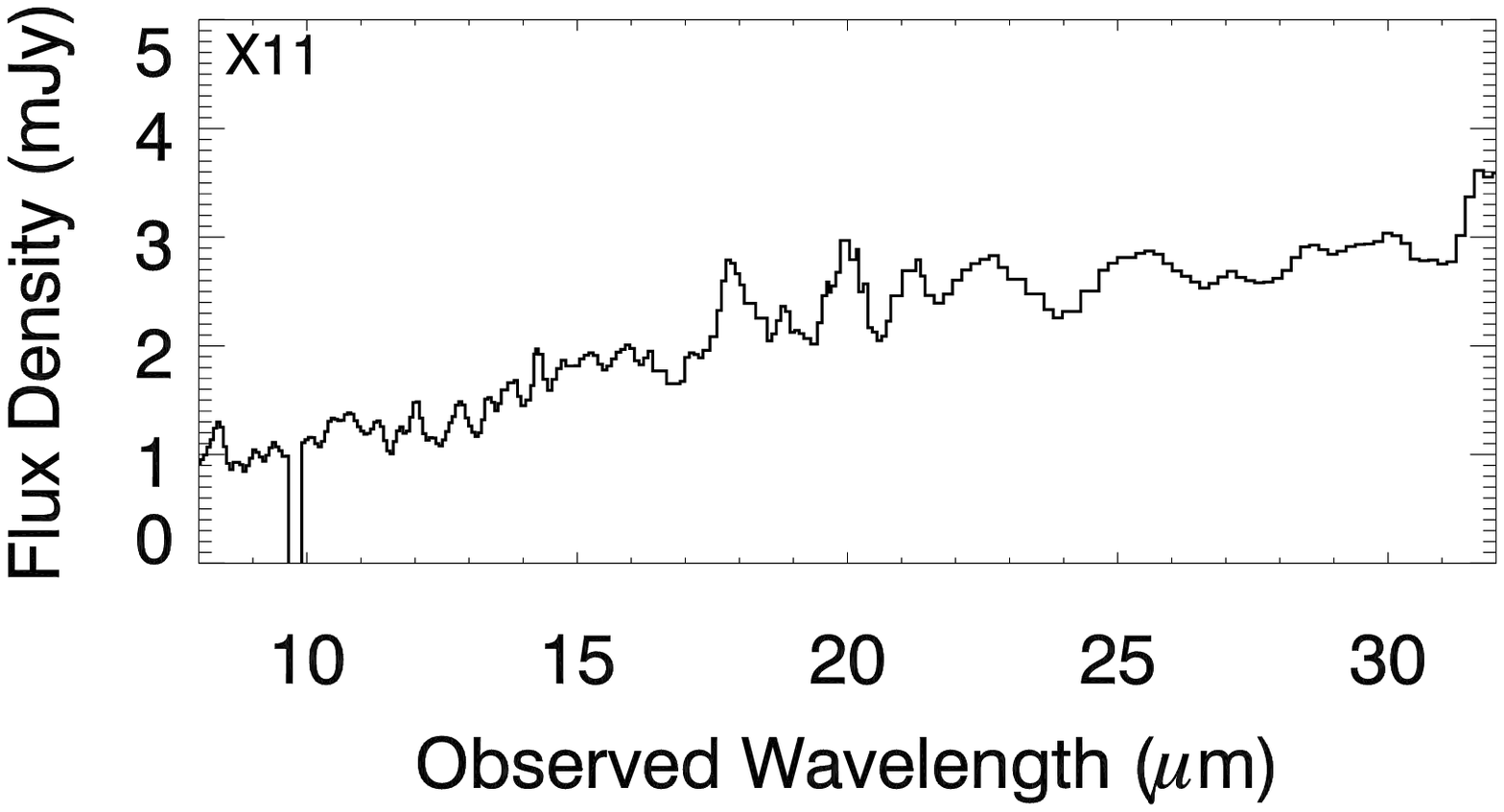}
\includegraphics[height=48mm]{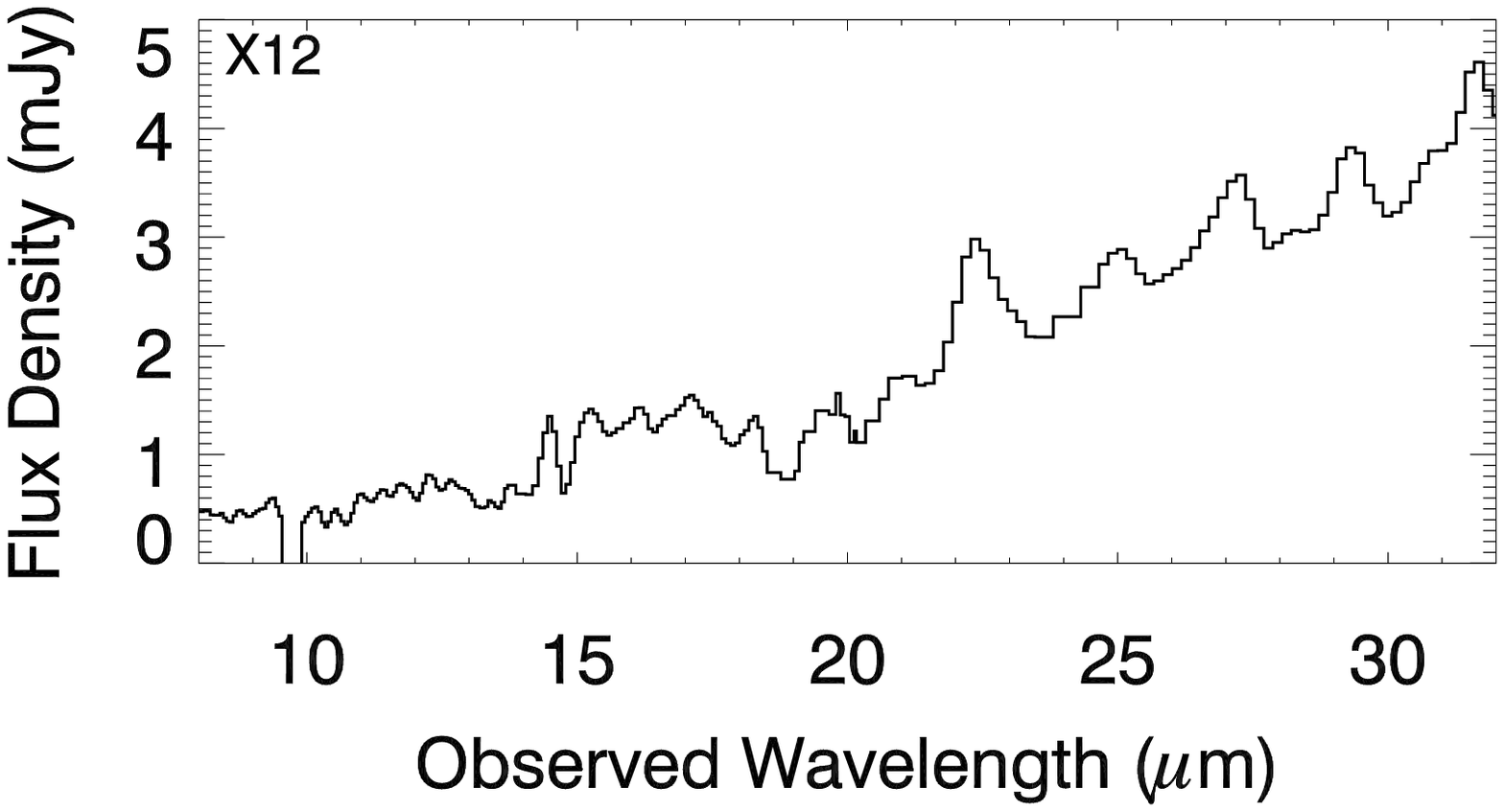}
\includegraphics[height=48mm]{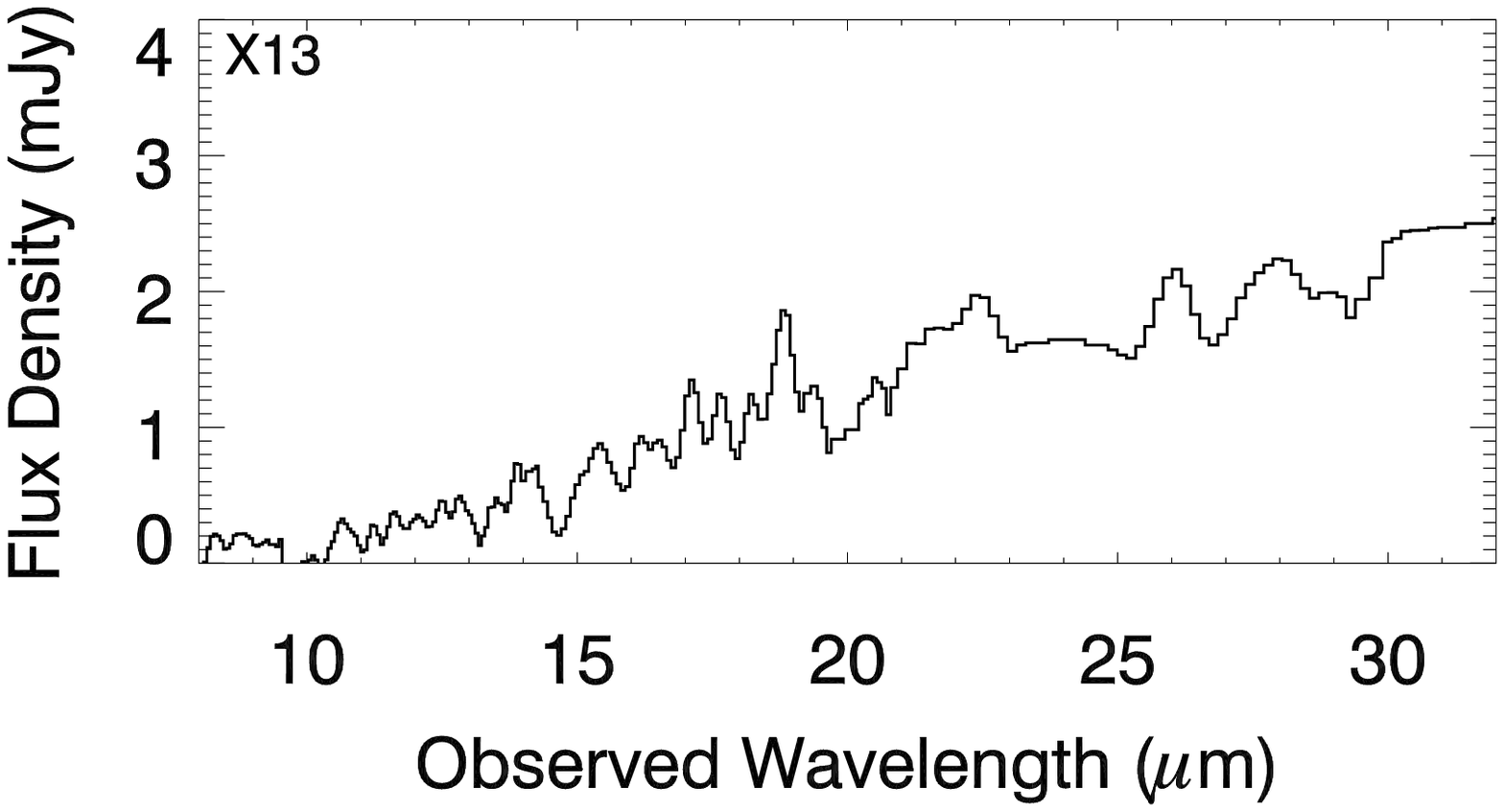}
\includegraphics[height=48mm]{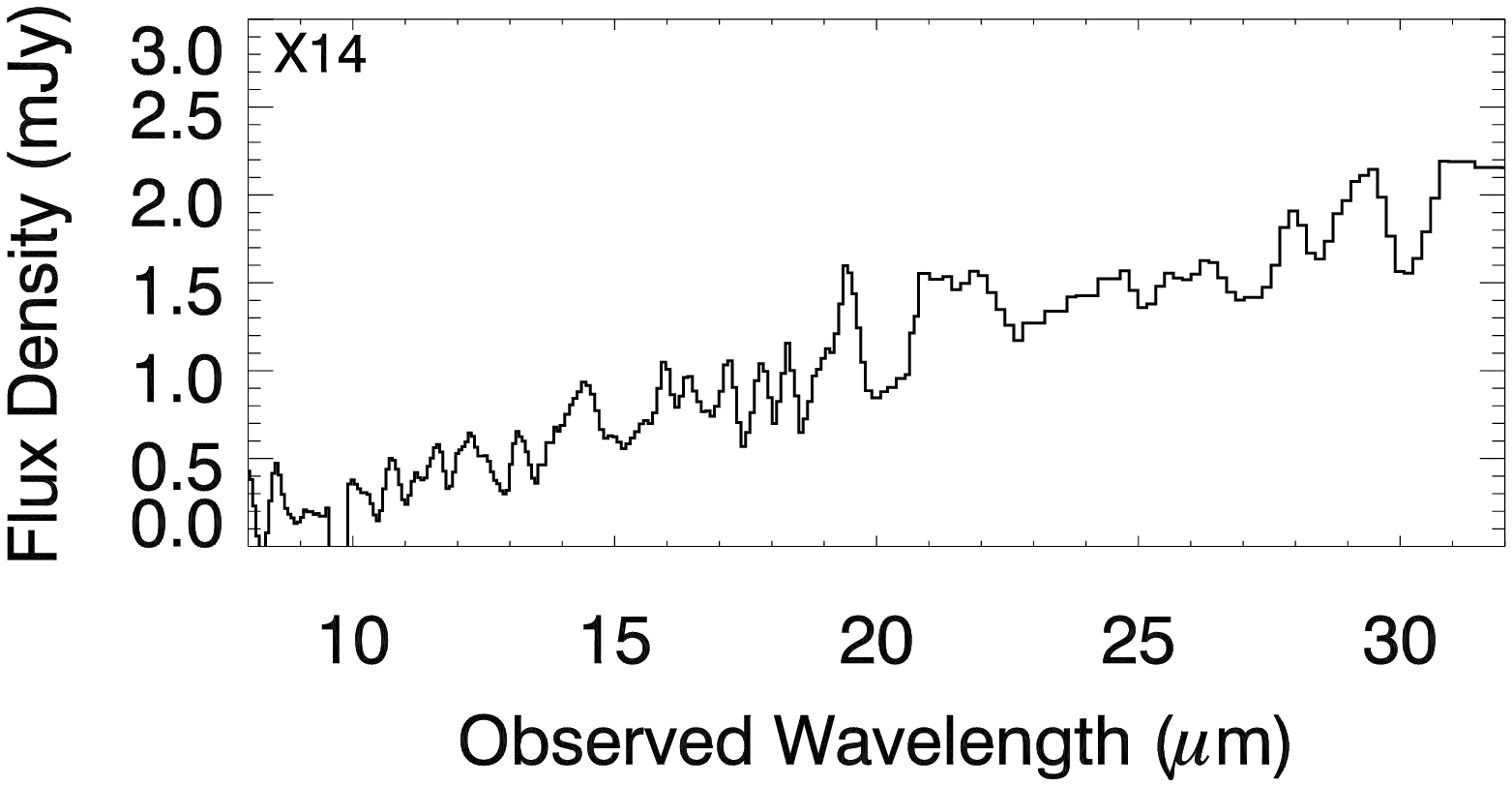}
\includegraphics[height=48mm]{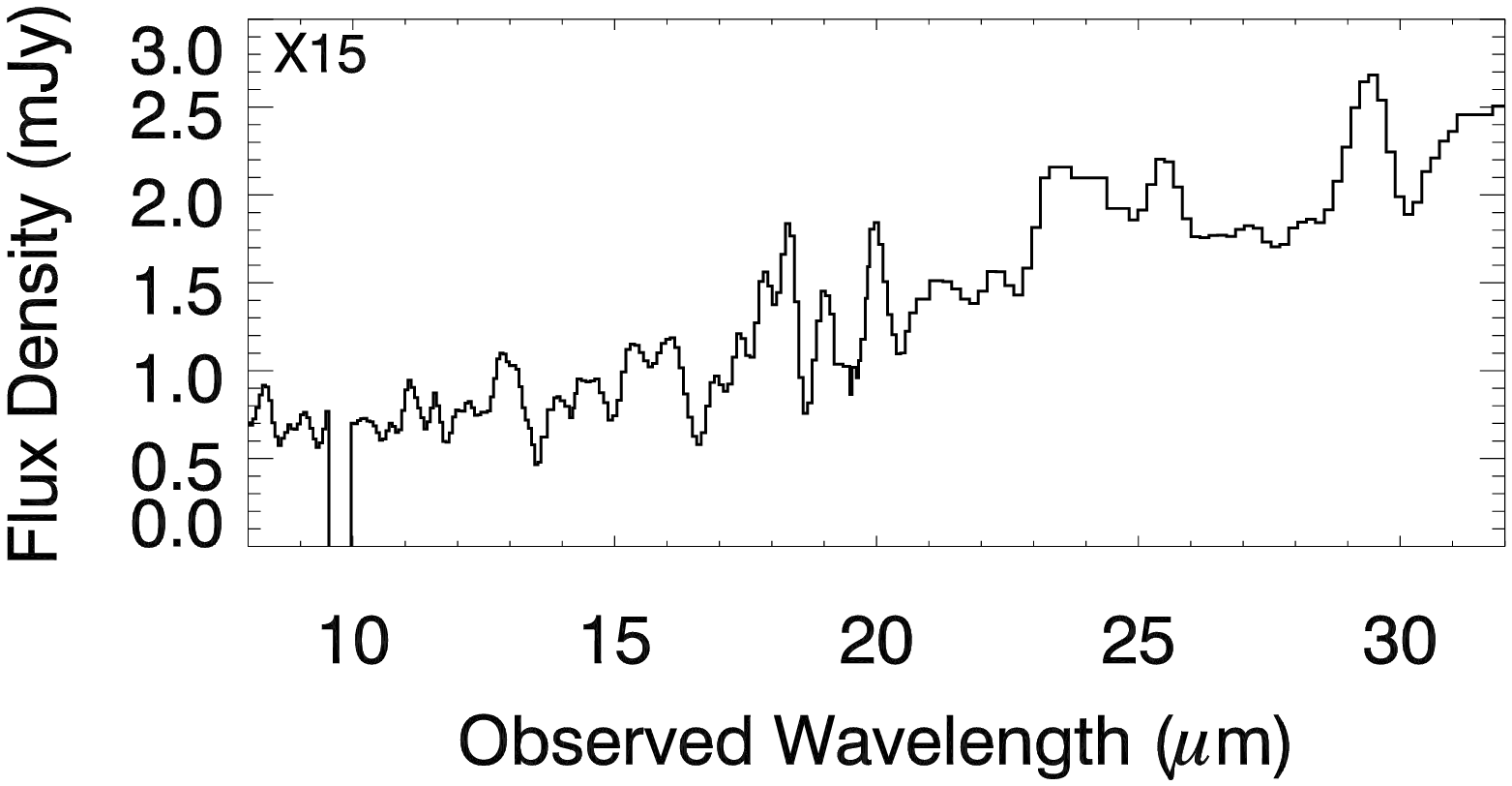}\\
{Fig. 1. --- Continued.}
\end{center}
\end{figure*}

\subsection{Absorption-Dominated Sources}

For the absorption-dominated sources, redshifts are determined primarily from the localized maximum in the IRS continuum, or ``hump'', which is blueward of the absorption feature. This feature is produced by absorption on either side of the feature, and with a possible contribution of 7.7$\mu$m PAH emission in composite sources.  We determine redshifts by assuming this hump has a rest-frame wavelength of 7.9$\mu$m (see average spectra in \citet{hao07} and \citet{spo07}). Comparison of IRS redshifts with optical redshifts for similar absorbed sources (e.g., \citealt{bra07}; \citealt{dey08}) suggests that they can be uncertain to $\pm$0.2 in $z$.

\begin{figure}[h]
\begin{center}
\includegraphics[height=60mm]{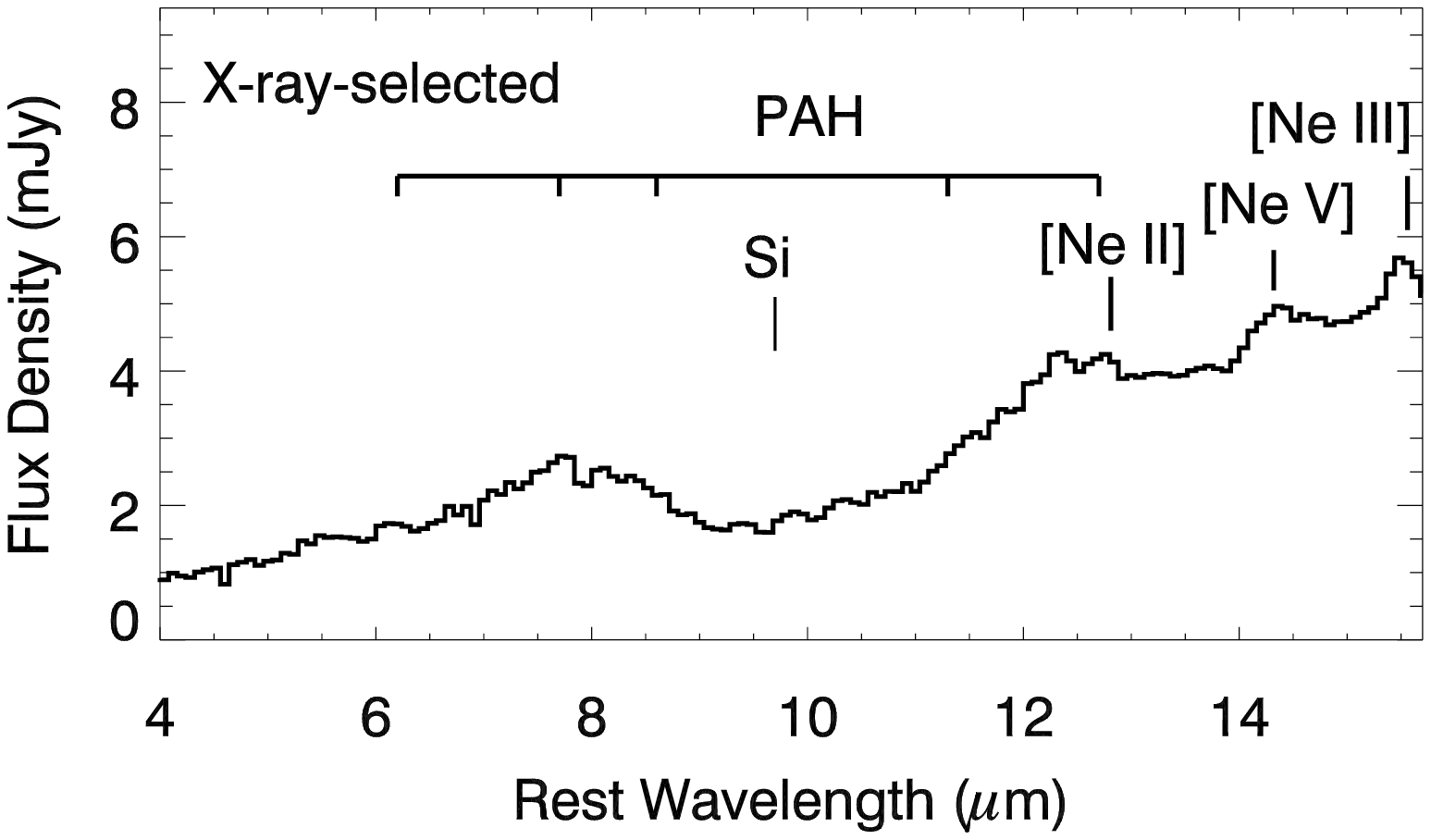}
\includegraphics[height=60mm]{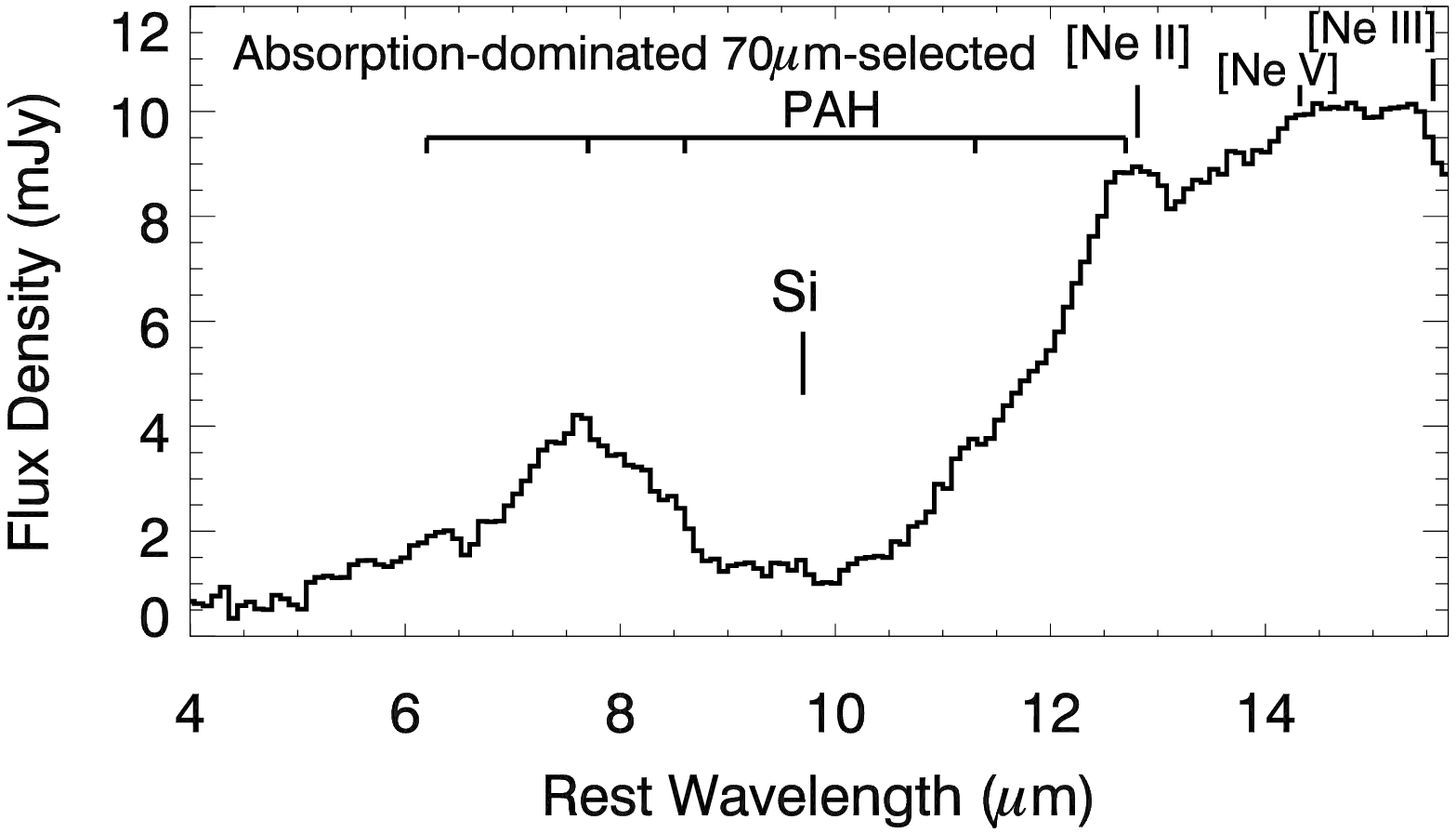}
\end{center}
\caption{\label{fig:xstack} Average IRS spectrum of X-ray sources with redshifts (top). For comparison, the average IRS spectra of absorption-dominated sources 70$\rm \mu m$ sources from \citet{bra08} is also shown (bottom). The expected positions of typically strong emission lines and the silicate absorption line are shown. The average spectrum combines the individual un-smoothed rest-frame spectra and is not smoothed.}
\end{figure}

For the absorption-dominated sources, the overall nature of the infrared spectra can be best considered by assembling the average IRS spectrum. This allows consistent features within individual spectra to be seen with improved signal-to-noise. This was achieved by wavelength correcting each individual unsmoothed spectrum to the rest-frame ($z$=0), interpolating them to a common wavelength scale of $\sim$0.1$\rm \mu m$/pix, and taking a straight (non-weighted) average at each wavelength position. The averaged spectrum is shown in Figure~\ref{fig:xstack}a. The most dominant feature is the silicate absorption feature at 9.7$\rm \mu m$. The average silicate absorption strength (as defined in Table~\ref{tab:irs}) is $S_{10}$=$-$1.0 (with individual values ranging from $-$0.4 to $-$2.9). For comparison, we also show the average spectrum of the absorption-dominated 70$\rm \mu m$-selected sources presented in \citet{bra08} (Figure~\ref{fig:xstack}b). Even though the X-ray-selected absorption-dominated sources have a similar average $R-$[24] color to that of the 70$\rm \mu m$-selected absorption-dominated sources, they have a much shallower average silicate absorption ($S_{10}$=$-$1.0 compared to $S_{10}$=$-$2.1). The average silicate absorption of the X-ray sample is substantially less than the median value for local ULIRGs (\citealt{hao07}; \citealt{spo07}) and the optically obscured sample observed by \citet{hou05} and \citet{wee06}, and is slightly larger that the median value for Seyfert 2 AGN ($-$0.6; \citealt{hao07}).

The individual spectra show no evidence for PAH emission (6.2$\rm \mu m$ equivalent width $<$0.2$\rm \mu m$; see Table~\ref{tab:irs}), although there is a hint of the 6.2 and 7.7$\rm \mu m$ PAH features in the average spectrum. However, because of the redshift uncertainties for the individual spectra, it is difficult to be confident in any weak narrow emission line features that appear in the average. The possible feature at 7.7$\rm \mu m$ could also be due to the [Ne VI] emission line at 7.6$\rm \mu m$. For each individual source, we use the noise spectrum to estimate the maximum flux from the 6.2$\rm \mu m$ PAH emission feature (we estimate the flux of the nearest noise feature to 6.2$\rm \mu m$). We then use the relation of \citet{pop07} to estimate the corresponding maximum SFR and infrared luminosity associated with star-formation. The values are shown in Table~\ref{tab:irs}. In some cases, the spectra are not sensitive enough to put a useful limit on the SFR. BootesX2 and BootesX7 have the lowest limits of SFR$<$190 M$_\odot$ yr$^{-1}$ and SFR$<$100 M$_\odot$ yr$^{-1}$ respectively, corresponding to infrared luminosities associated with star-formation of log L$_{IR}<$12.0 and $<$11.7 respectively. Comparing these values to the total infrared luminosity estimated from the rest-frame 8$\rm \mu m$ flux density, suggests that less than 30\% and 13\% of the total infrared luminosities are from star formation processes. 

Weak indications are present in the average for [Ne II], [Ne V], and [Ne III] emission lines. The relative intensities of these lines are $\sim$0.7:1:1. The presence of [Ne V] is a strong AGN indicator (\citealt{lut98}; \citealt{far07}). The relative intensities show that [Ne V] is stronger than that of local Seyferts such as NGC 4151, Markarian 3, and CenA \citep{wee05}. If there is a contribution of 12.7$\rm \mu m$ PAH emission to the [Ne II] line in the stacked spectrum, this would make this effect even larger. This may indicate a higher ionization of the narrow-line region by the more powerful AGN in these sources. 

We have measured the luminosities at 5.8$\rm \mu m$ to determine the continuum luminosity just shortward of the 6.2$\mu$m PAH feature and at a wavelength without strong absorption (e.g., \citealt{brl06}; \citealt{saj07}; \citealt{hao07}: see Table~\ref{tab:irs} for individual values). The absorption dominated sources have median luminosity $\nu$L$_{\nu}$(6$\mu$m) = $6 \times 10^{45}~\rm ergs~s^{-1}$ (1.5 $\times 10^{12}$ L$_{\odot}$). These are very luminous sources, exceeding that of most local AGN in the sample of \citet{hao07}: Sy 2 in that sample have median $\nu$L$_{\nu}$(6$\mu$m)$\sim 10^{10}$L$_{\odot}$, and ULIRGs have $\nu$L$_{\nu}$(6$\mu$m)$\sim$ $10^{11}$L$_{\odot}$. \citet{des07} find that local ULIRGs span the range $\nu$L$_{\nu}$(6$\mu$m) = 1$ \times 10^{10}-5\times 10^{12}$ L$_\odot$. Our sources lie at the upper end of this range. The most luminous X-ray source, at $\nu$L$_{\nu}$(6$\mu$m)=4 $\times$ 10$^{12}$ L$_\odot$, is nearly as luminous as the most luminous source yet published from IRS spectra: source 9 in \citet{hou05} with $\nu$L$_{\nu}$(6$\mu$m)=5.9 $\times$ 10$^{12}$ L$_\odot$.

We estimate the total infrared luminosity, L$_{IR}$=L(8-1000$\rm \mu m$) of each of the absorption dominated sources. The conversion factors vary significantly depending on the template adopted for the conversion \citep{cap07}. Because we expect the infrared emission to be dominated by obscured AGN activity, we use the conversion for the template Mrk 231: L$_{IR}$=5$\times \nu$L$_{\nu}$(8$\mu$m), where the rest-frame 8$\rm \mu m$ flux density is measured directly from the IRS spectra. The estimates are about a factor of 7 lower than one would obtain using templates of starburst-dominated galaxies such as Arp 220, and this explains our smaller L$_{IR}$ compared to that of  \citet{hou05} for Houck13 which was estimated from F00183-7111. The infrared luminosities confirm that the absorption-dominated sources are all ultra-luminous ($\rm L_{IR}>$10$^{12}~\rm L_{\odot}$) or hyper-luminous ($\rm L_{IR}>$10$^{13}~\rm L_{\odot}$) infrared galaxies.

\subsection{Power-law Sources}
\label{sec:pow}

Given the lack of features and/or optical follow-up of the power-law sources, we do not have redshifts for these sources. We therefore cannot construct an average spectrum or determine luminosities. The spectra of the power-law sources must have weak or absent silicate features (6.2$\rm \mu m$ PAH equivalent widths $\ltsimeq$0.2$\rm \mu m$) or they lie beyond $z\sim$2.6, causing their absorption features to fall longward of the IRS wavelength limit. It is difficult to determine whether silicate emission exists in these sources given our lack of redshift information and that the silicate emission feature is generally weaker than absorption feature. However BootesX6 may have a weak silicate emission feature if it is at $z$=0.9. All 3 of the ``X-ray bright'' subsample are power-law sources. 

\section{Discussion}

\subsection{IRS spectra}

All the X-ray selected sources have infrared spectra consistent with their mid-IR luminosities being dominated by emission originating from AGN activity. This is expected given that their high X-ray luminosities imply that they host luminous AGN which should dominate their infrared spectra.  We find a larger fraction of power-law sources (4/13 of the main sample and 7/16 of the total (main + ``X-ray bright'') sample) than is found in the SWIRE sample of \citet{wee06b} (they find only 1/9 to have a featureless power-law spectrum in their sample which has an X-ray limit which is $\approx 8 \times$ fainter than that of our sample). This may in part be due to them having accurate optical redshifts, which make it easier to identify features. It could also be due to our selection of more luminous X-ray sources, which are less likely to be heavily obscured and exhibit silicate absorption. This hypothesis is supported by the fact that 6/7 of the power-law sources are also the brightest X-ray sources in the sample. 

To explain the lack of a silicate absorption feature in their IRS spectra, the power-law sources must have weak or absent silicate features or they must lie beyond $z\sim$2.6. We argue that the latter is unlikely given that they are among the brightest X-ray sources in our sample. If at $z>$2.6, they must be extremely intrinsically luminous. Assuming a canonical power-law spectrum with photon index, $\Gamma$=1.7, the typical X-ray flux of f$_{0.5-7 keV}$=4 $\times 10^{-14}\rm~ergs~cm^{-2}s^{-1}$ corresponds to a rest-frame X-ray luminosity of L$_{0.5-7 keV}$=2 $\times 10^{45}\rm~ergs~s^{-1}$ at $z$=2.6 which would suggest that they were among the most powerful quasars known. If at $z>2.6$, these sources would be significantly more X-ray luminous than any of the sources at $z<2.6$. We would not expect a sharp difference in the luminosities of sources at higher redshifts. Furthermore, only 0.7\% of the $\sim$1000 X-ray sources in the $Chandra$ XBo\"otes survey \citep{ken05} with spectroscopic redshifts from AGES (Kochanek et al. in prep) have luminosities this high. In the following discussion, we assume that these sources are at similar redshifts to the rest of the X-ray selected sources and that they must have weak or absent silicate features.  

\subsection{Infrared colors of X-ray selected AGN}

\begin{figure*}[h]
\begin{center}
\includegraphics[height=70mm]{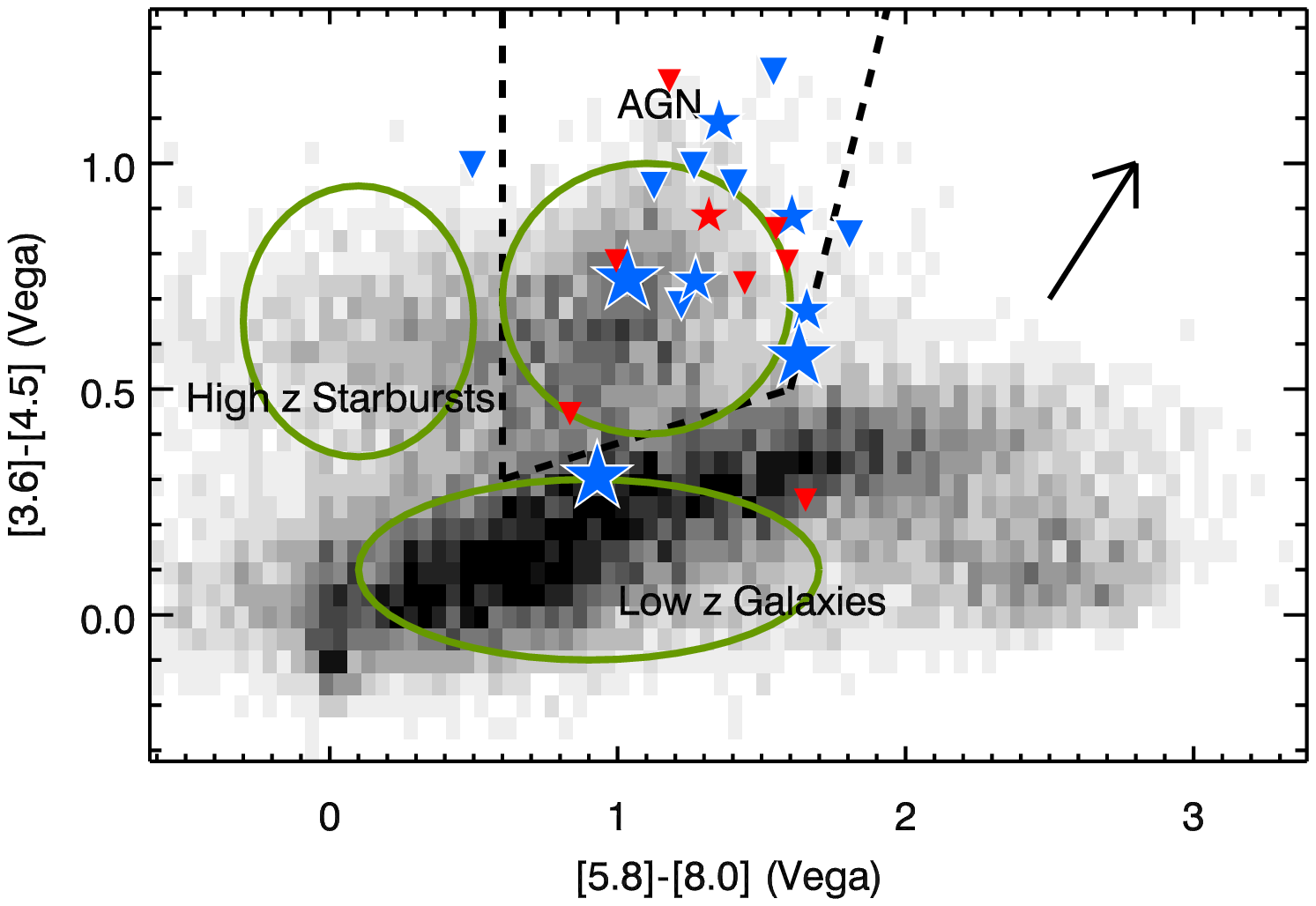}
\end{center}
{\caption[junk]{\label{fig:irac} IRAC color-color diagram for X-ray selected IRS sources (large symbols). The Bo\"otes X-ray selected sources are denoted by large blue symbols (with the largest symbols denoting the X-ray bright sub-sample). The fainter X-ray sources from \citet{wee06b} are also shown (smaller red symbols). The sources are split into silicate absorption-dominated sources (downwards triangles), and power-law dominated sources (stars). We show the distribution of all $\approx$10,000 MIPS sources with f$_{\nu}$(24$\rm \mu m$)$>$0.5 mJy in greyscale. The rough regions expected to be inhabited by different populations are also shown. The black dashed line is that proposed by \citet{ste05} to empirically separate AGN from Galactic stars and normal galaxies. The arrow represents the reddening vector, estimated using the extinction curve of \citet{dra03}, for a source at $z$=1 reddened by a dust column corresponding to R$_v$=3.1.}}
\end{figure*}

In Figure~\ref{fig:irac}, we show the IRAC color-color diagram for our IRS sources. To increase the number of X-ray selected sources, we also plot the 9 X-ray selected IRS sources from the Lockman Hole field of the SWIRE survey presented by \citet{wee06b}. The SWIRE sample is selected to have f$_{\nu}$(24$\rm \mu m$)$>$0.9 mJy, f$_{0.3-8 keV} > 10^{-15} \rm ~ergs~cm^{-2}~s^{-1}$, and $R>22$ mag. The X-ray flux limit still corresponds to typical AGN luminosities at these redshifts. To determine the location in color-color space of these sources in comparison to the general 24 $\rm \mu m$ population, we also plot the distribution of the $\approx$10,000 MIPS Bo\"otes sources with f$_{\nu}$(24$\rm \mu m$)$>$0.5 mJy. The rough regions expected to be inhabited by different populations are shown. We also plot the empirical line of \citet{ste05} (see also \citealt{lac04}) to separate AGN-dominated sources from Galactic stars and normal galaxies. 

20/25 of the X-ray selected sources lie within the AGN region of \citet{ste05}. \citet{hic07} show that large amounts of extinction (A$_v>$50) toward the active nucleus and/or a large contribution from a starburst can cause the IRAC colors of an object to fall outside of the \citet{ste05} region. Because our sources are largely within the AGN region, they are unlikely to be heavily extincted or have a large starburst contribution to their mid-IR emission. There appears to be no correlation between the position in IRAC color-color space with the presence of the silicate absorption strength, again supporting the hypothesis that the silicate feature arises in the cooler gas which is traced by longer infrared wavelengths. The sources in the X-ray bright sub-sample appear to have slightly bluer colors which may demonstrate that they are less dust obscured. The IRAC color-color classification is expected to deteriorate at $z\gtsimeq$1 as the observed IRAC bands are pushed towards lower wavelengths in which the stellar light dominates over hot dust emission (\citealt{ste05}; \citealt{lac04}). However, it likely holds to higher redshifts for the most luminous sources because the near-infrared emission is still dominated by the hot dust, even in the integrated spectral energy distribution. 
\subsection{Infrared-to-optical colors of X-ray selected AGN}

\begin{figure*}[h]
\begin{center}
\includegraphics[height=70mm]{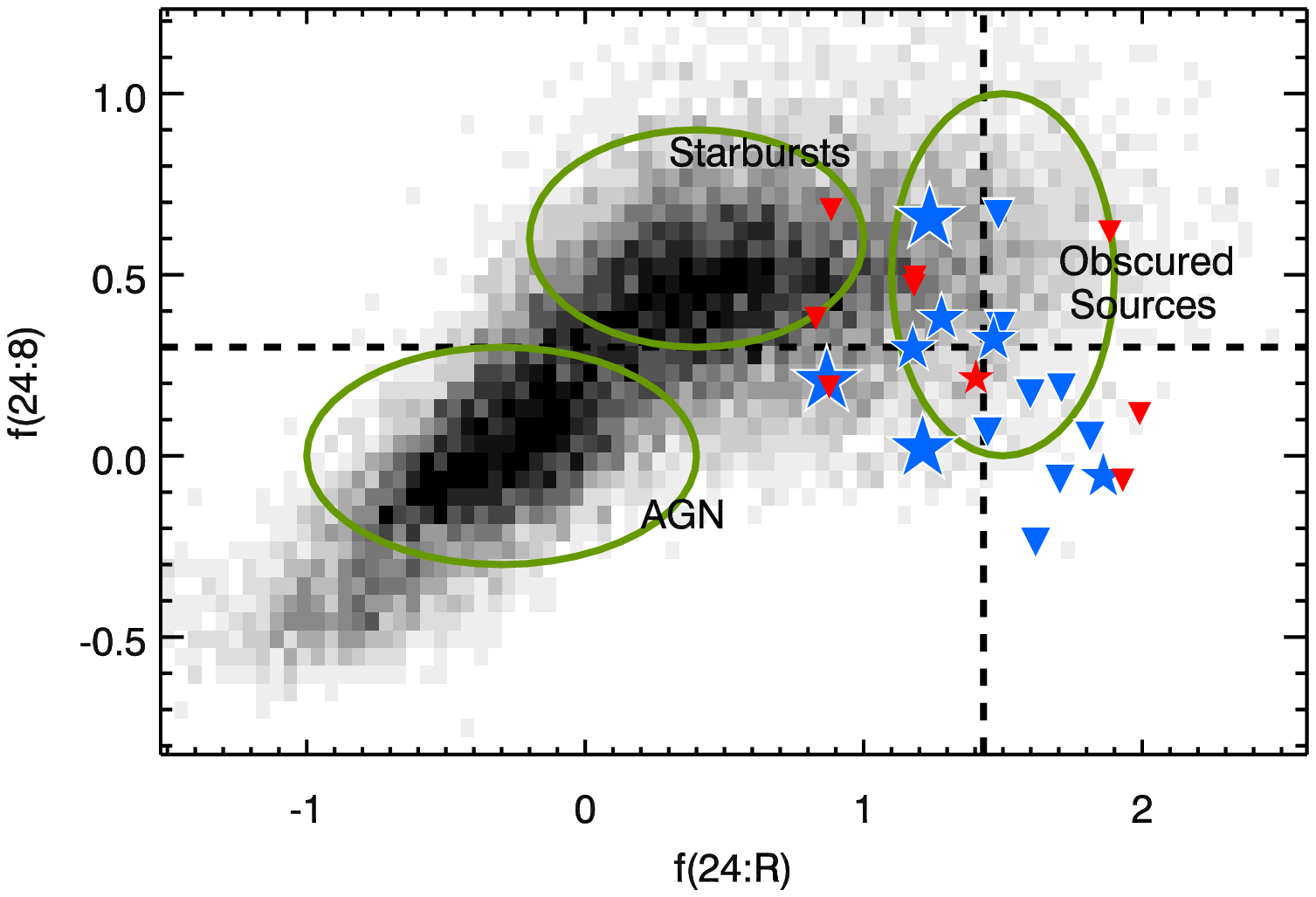}
\end{center}
{\caption[junk]{\label{fig:opt_col} 24-to-8 $\rm \mu m$ vs. 24-to-0.7 $\rm \mu m$ color-color diagram for X-ray selected IRS sources. The 24-to-8 $\rm \mu m$ and  24-to-0.7 $\rm \mu m$ colors are defined as f(24:8)=$\rm log_{10}$[$\rm \nu f_\nu(24 \mu m)$/$\rm \nu f_\nu(8 \mu m)$] and f(24:R)=$\rm log_{10}$[$\rm \nu f_\nu(24 \mu m)$/$\rm \nu f_\nu(0.7 \mu m)$] respectively. The symbols and colors are the same as in Figure~\ref{fig:irac}. The vertical dashed line shows the R-[24]$>$14 (log($\rm \nu f_{\nu}$(24)/$\rm \nu f_{\nu}$(R)$>$1.43) criteria used to select the powerful obscured sources presented in \citet{hou05} and \citet{wee06}. The horizontal dashed line shows the 24-to-8$\rm \mu m$ color criteria used by \citet{bra06b} to roughly divide steeper spectrum starburst sources from shallow spectrum AGN-dominated sources at $z>$0.6.}}
\end{figure*}

In Figure~\ref{fig:opt_col}, we show the 24-to-8 $\rm \mu m$ vs. 24-to-0.7 $\rm \mu m$ color-color diagram for our sources (see \citet{yan04} for the location of different populations in this color-color space). The 24-to-8 $\rm \mu m$ and  24-to-0.7 $\rm \mu m$ colors are defined as f(24:8)=$\rm log_{10}$[$\rm \nu f_\nu(24 \mu m)$/$\rm \nu f_\nu(8 \mu m)$] and f(24:R)=$\rm log_{10}$[$\rm \nu f_\nu(24 \mu m)$/$\rm \nu f_\nu(0.7 \mu m)$] respectively. f(24:R) is a good indicator of obscuration and/or redshift. f(24:8) has been used as a crude measure of the spectral slope and hence the dust temperature distribution and whether the source is AGN- or starburst-dominated. \citet{bra06b} find that AGN-dominated sources tend to have f(24:8)$<$0.3 and starburst-dominated sources tend to have f(24:8)$>$0.3. Even though the X-ray selected sources must harbor a powerful AGN, their 24-to-8 $\rm \mu m$ colors suggest that the sources are dominated by a combination of AGN and starbursts. We suggest that this diagnostic breaks down for obscured sources due to the heavy absorption at near-IR wavelengths which will steepen the infrared slope regardless of the underlying power source. In addition, at $z\approx$1-2, the silicate absorption feature is moving through the 24$\rm mu m$ filter, complicating the interpretation further.  

The $R$-[24] colors of the X-ray selected sources are typically bluer than that of the most extreme IRS sources presented in \citet{hou05} and \citet{wee06} which typically have $R-[24]>$15 (f(24:R)$>$1.83). The faintest optical magnitude is relatively bright ($R$=23.7) despite the fact that there is no optical selection criteria against including optically faint sources. We suggest that the X-ray-selection is biasing us against the most heavily absorbed (and therefore optically faintest) sources. Indeed, the sources in the X-ray bright sub-sample have among the bluest $R$-[24] colors due at least partially to their relatively bright $R$-band magnitudes. 

\subsection{Correlations between Infrared and X-ray properties}

\begin{figure}[h]
\begin{center}
\includegraphics[height=50mm]{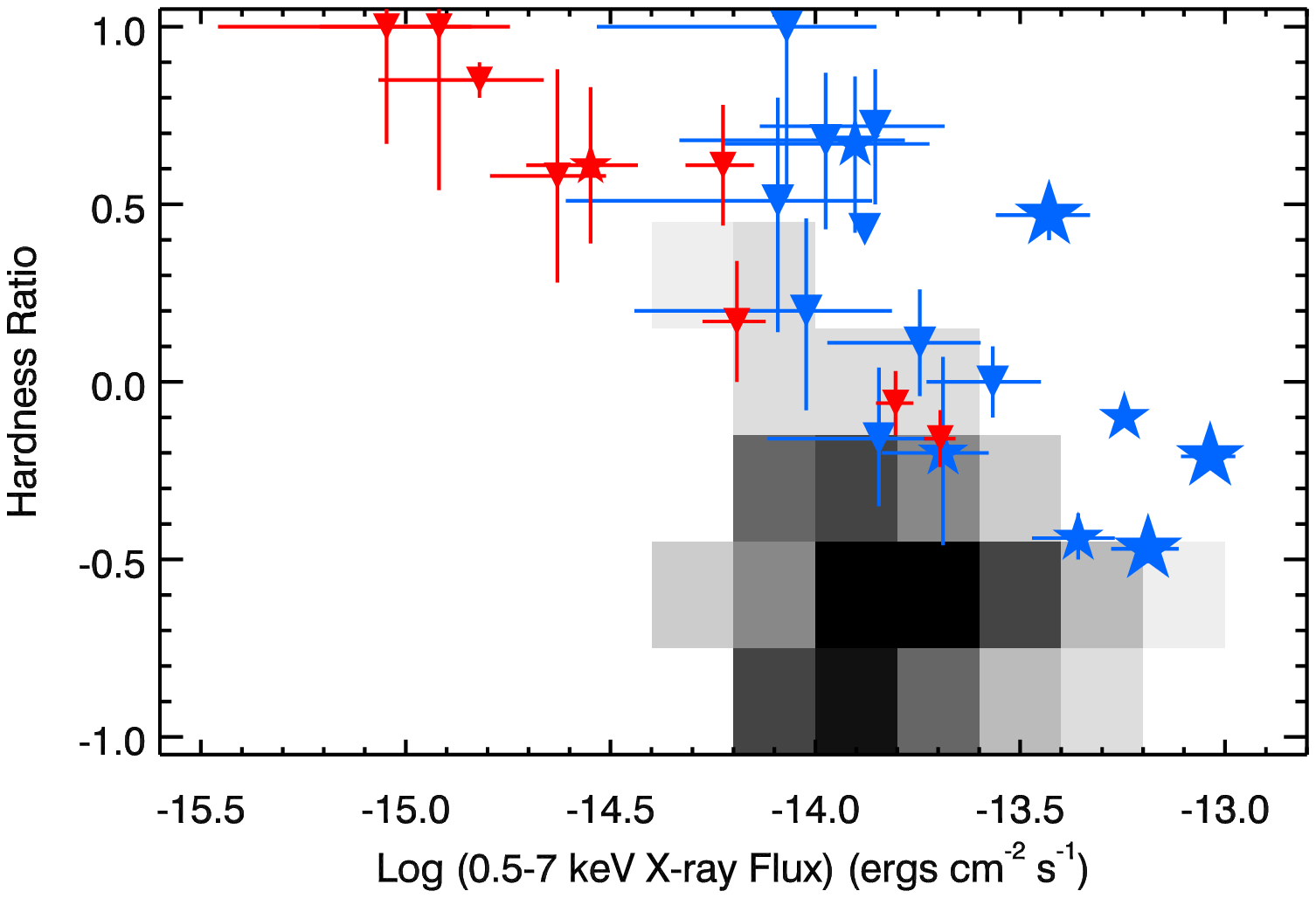}
\end{center}
{\caption[junk]{\label{fig:xabs} 0.5-7 keV band X-ray flux against X-ray hardness ratio for X-ray selected IRS sources. The hardness ratios can range from 1 (hardest X-ray spectrum) to -1 (softest X-ray spectrum). The Bo\"otes X-ray selected sources are denoted by large blue symbols (with the largest symbols denoting the X-ray bright sub-sample). The fainter X-ray sources from \citet{wee06b} are also shown (smaller red symbols). The sources are split into silicate absorption-dominated sources (downwards triangles), and power-law dominated sources (stars). Poisson error bars are also shown. The grey scale shows the distribution of all X-ray sources in the XBo\"otes survey with redshifts $z$=0.9-2.6.}}
\end{figure}

Figure~\ref{fig:xabs} shows the observed X-ray flux as a function of X-ray hardness ratio for the combined sample of Bo\"otes and SWIRE X-ray IRS sources. There are no sources with both small X-ray fluxes and small X-ray hardness ratios. The grey-scale shows that this is not seen in the larger population of X-ray sources in the same redshift range, and only becomes apparent when we jointly select X-ray sources that are both infrared bright and optically faint. Our selection criteria result in a sample of bolometrically luminous high redshift AGN. In these cases, small observed X-ray fluxes will be associated with more obscured sources which will also have larger hardness ratios.

For our combined sample, the absorption-dominated sources tend to have fainter X-ray fluxes and larger X-ray hardness ratios (with a median value of HR=0.55) which are indicative of more absorbed X-ray spectra and larger HI column densities. Conversely, the power-law sources have brighter X-ray fluxes and smaller hardness ratios (with a median value of HR=$-$0.2) indicative of less obscured X-ray spectra and smaller HI column densities. This suggests that the dusty absorbing regions responsible for the silicate absorption also contain sufficient gas to cause X-ray absorption or that both the gas and dust lie along similar lines of sight. Deeper silicate absorption features may be associated with more embedded AGN with higher column densities of gas. Our results are consistent with those of \citet{shi06} who measure the silicate absorption depth for a sample of 85 nearby AGN with X-ray data and find a correlation between the strength of the silicate feature and the HI column density. Our lack of redshifts for the power-law sources means that we must compare the X-ray fluxes and hardness ratios rather than the X-ray luminosities and fitted HI column densities. Any large difference in the redshift distribution of the two populations could affect our results (for example higher redshift sources with identical X-ray luminosities will have softer X-ray spectra and smaller X-ray fluxes). However, in Section~\ref{sec:pow}, we argue that the power-law sources lie at $z<$2.6 and there is no reason to think that they lie at significantly different redshifts to the absorption-dominated sources given their similar selection criteria. 
 
Although we find a trend in the X-ray and infrared properties of these sources, the correlations are only weak. The difficulty in establishing a clear relation between X-ray and infrared properties may be explained by the distribution of their gas and dust. \citet{ris02} show that the X-ray absorbing column density of Seyfert 2 galaxies varies on yearly timescales for most sources. This suggests that most of the X-ray absorption must occur on sub-parsec scales. They obtain even tighter constraints in \citet{ris07} for a Seyfert galaxy which exhibits spectral variation on timescales of days. This suggests that, at least for less powerful sources, the obscuring gas is on a similar scale to that of the broad-line region. Significant amounts of dust are unlikely to lie at these distances because of the dust sublimation temperatures. The difference in the location of much of the gas and dust may explain the large scatter between parameters tracing the X-ray and mid-IR obscuration (e.g., \citealt{rig04}). \citet{lev07} suggest that the obscuring material in dusty AGN is likely to consist of both dense clumpy clouds of dust and a smoother inter-cloud medium. They argue that deep silicate absorption can only be produced in a source embedded in a smooth medium since clumpy clouds will provide direct views of portions of illuminated clouds which will fill in the resultant silicate absorption feature. Since our X-ray selected sources have shallower silicate absorption features than typical ULIRGs, this may suggest that they are dominated by clumpy obscuration. Alternatively, they may simply have a smaller line of sight obscuration. 

\citet{stu06} obtained IRS spectra for a sample of 7 X-ray selected AGN which were specifically chosen to have large X-ray luminosities and large column densities of gas (due to their large X-ray hardness ratios). Given the correlation between HI column density and silicate absorption strength (see discussion above and \citealt{shi06}), we might expect them to exhibit deep silicate features. However, all of their sources exhibit featureless power-law spectra. The solution to this apparent problem may be that they selected a highly unusual population of sources which were both extremely X-ray luminous and X-ray obscured. Figure~\ref{fig:xabs} shows that although X-ray obscured sources are more likely to have silicate absorption features, X-ray luminous sources are likely to have featureless power-law spectra. Perhaps more luminous AGN are more capable of breaking up the surrounding dust into discrete clouds which would fill in the silicate absorption feature in these cases despite the large column densities of gas. 

We compare the fraction of power-law dominated to absorption dominated sources in different samples. \citet{wee06b} probe the lowest X-ray luminosities and find only 1/9 (11\%) are power-law dominated. Our sample has higher X-ray luminosities and we find the fraction of power-law to absorption dominated sources to be 4/13 (31\%) in the main sample and 7/16 (43\%) in the total (main + X-ray bright) sample. All of the sources in the \citet{stu06} study are power-law dominated. Their sources are particularly bright in the X-ray, with intrinsic X-ray luminosities in the range logL$_x$=44.0-45.5. This suggests that more X-ray luminous sources are more likely to have shallower silicate absorption (which will appear as power-law sources).

\begin{figure}[h]
\begin{center}
\includegraphics[height=50mm]{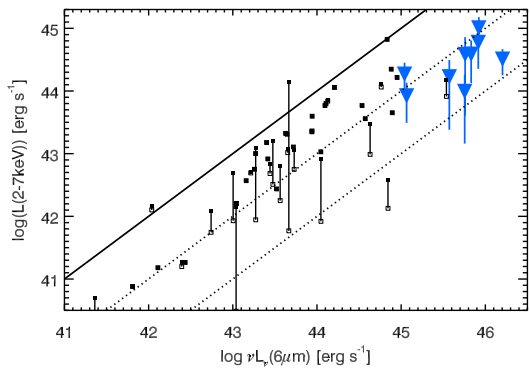}
\end{center}
{\caption[junk]{\label{fig:xvs6} Rest-frame 6$\rm \mu m$ luminosity against rest-frame absorption-uncorrected 2-7 KeV X-ray luminosity for absorption-dominated Bo\"otes sources (blue triangles). The Poisson error bars are shown for the X-ray luminosity. The low redshift ISO sources from \citet{lut04} are plotted as small squares (connected empty and filled squares denote X-ray luminosities uncorrected and corrected for X-ray absorption respectively). The solid line shows a direct correspondence between the X-ray and 6$\rm \mu m$ luminosities. The dotted lines show where L(2-7keV)=0.1 $\nu L_\nu$(6$\rm \mu m)$ and L(2-7keV)=0.01 $\nu L_\nu$(6$\rm \mu m)$.}}
\end{figure}

\citet{kra01}, \citet{lut04}, and \citet{hor06} show that there is a strong relation between AGN hard X-ray emission and mid-IR continuum in low redshift active galaxies. Figure~\ref{fig:xvs6} shows the AGN continuum luminosity at 6$\rm \mu m$ as a function of 2-7 keV X-ray luminosity for the absorption-dominated sources (the power-law sources could not be included because we do not have redshifts to derive the luminosities). Although the absorption-dominated sources do not show a clear correlation between these two quantities, this lack is likely to be due to the large luminosity errors and relatively narrow luminosity range of the sample. When we compare our sources with lower luminosity sources at low redshift \citep{lut04}, we find that they form the higher luminosity extension of the relation. We have chosen not to correct the X-ray luminosity for absorption given the large uncertainties. 7/9 of our sources lie at L(2-7keV)/$\nu$L$_\nu$(6$\rm \mu m$)$<$0.1, compared to only 8/47 of the \citet{lut04} sources (before correcting for absorption). This suggests that our sources are generally more obscured than the low redshift, lower luminosity AGN presented in \citet{lut04}.

\section{Conclusions}

We have presented mid-infrared spectra obtained using $Spitzer$ IRS of a sample of 16 X-ray loud 24 $\rm \mu m$ sources. The spectra of these sources are characterized by either silicate absorption or a featureless power-law, with no evidence of PAH emission. This suggests that the mid-infrared emission from these sources is dominated by warm dust emission surrounding the active nucleus with little contribution from star formation. The absorption-dominated sources lie in the range 0.9$<z<$2.6 and have very high X-ray and infrared luminosities (log L$_{IR}$=12.5-13.6 L$_\odot$). The average absorption depth in our absorption-dominated X-ray sources is similar to that of local Seyfert 2 AGN and much smaller than that of typical ULIRGs. This suggests a difference in the distribution of the dusty material between our sample and typical ULIRGs. The weak silicate absorption feature in our sample could indicate a more clumpy dust distribution or a lower line of sight obscuration. There is evidence in the average spectrum for weak [Ne II], [Ne V], and [NeIII] lines. The [Ne V] is relatively stronger than is seen in local Seyferts, suggesting a higher ionization of the narrow-line region in these more powerful AGN. Given their higher X-ray fluxes, it is unlikely that the power-law sources lie at $z>$2.6. Instead, they must have only weak silicate absorption or emission features. Comparison with other samples suggests that the fraction of power-law sources appear to increase with X-ray luminosity and decrease with X-ray obscuration. Both the infrared and infrared to optical colors suggest that the sources are relatively unobscured compared with $R$-[24]$>$15 sources. Our X-ray selection is likely to be preferentially biased to less obscured and/or more powerful AGN. We find that sources with silicate absorption features tend to have fainter X-ray fluxes and larger hardness ratios. This suggests that more absorbed sources also have higher column densities of absorbing gas along the same line of sight.

\acknowledgements
We thank our colleagues on the NDWFS, XBo\"otes, MIPS, IRS, AGES, and IRAC teams. KB is supported by the Giacconi fellowship at STScI. This research is supported by the National Optical Astronomy Observatory which is operated by the Association of Universities for Research in Astronomy, Inc. (AURA) under a cooperative agreement with the National Science Foundation. Support for this work by the IRS GTO team at Cornell University was provided by NASA through Contract Number 1257184 issued by JPL/Caltech. This work is based on observations made with the {\it Spitzer Space Telescope}, which is operated by the Jet Propulsion Laboratory, California Institute of Technology under a contract with NASA. The {\it Spitzer / MIPS} survey of the Bo\"otes region was obtained using GTO time provided by the {\it Spitzer} Infrared Spectrograph Team (James Houck, P.I.) and by M. Rieke. The IRS was a collaborative venture between Cornell University and Ball Aerospace Corporation funded by NASA through the Jet Propulsion Laboratory and the Ames Research Center. We thank the anonymous referee for his/her useful comments.


\end{document}